\documentclass[acmsmall, screen]{acmart}
\AtBeginDocument{%
  }

\acmYear{2026}
\acmDOI{XXXXXXX.XXXXXXX}
\acmISBN{978-1-4503-XXXX-X/2018/06}

\usepackage{tabularx, float, booktabs, adjustbox, multirow, makecell}
\usepackage{pifont}
\usepackage[linesnumbered,ruled,vlined]{algorithm2e}
\usepackage{graphicx}
\usepackage{comment}
\usepackage{textcomp}
\usepackage[table]{xcolor}
\definecolor{lightgray}{gray}{0.92}
\definecolor{lightblue}{HTML}{f0f0f5} 
\definecolor{lightgr}{HTML}{e0ebeb} 
\usepackage{listings}

\usepackage{subcaption}

\definecolor{keywordblue}{rgb}{0.26, 0.13, 0.86} 

\definecolor{codegreen}{rgb}{0,0.6,0}
\definecolor{codegray}{rgb}{0.5,0.5,0.5}
\definecolor{codepurple}{rgb}{0.58,0,0.82}
\definecolor{backcolour}{rgb}{0.95,0.95,0.92}

\definecolor{pragmagreen}{rgb}{0.0, 0.5, 0.0}  

\usepackage{enumitem}

\usepackage{tikz}

\usepackage{float}
\floatstyle{plain}
\newfloat{Code}{tbp}{lop}
\floatname{Code}{Listing}

\begin{document}

\title{ARMOR: Robust and Efficient CNN-Based S\underline{AR} ATR through \underline{Mo}del-Ha\underline{r}dware Co-Design}

\author{Sachini Wickramasinghe}
\authornote{Both authors contributed equally to this research.}
\orcid{0009-0000-8731-8166}
\email{shwickra@usc.edu}
\affiliation{%
  \institution{University of Southern California}
  \city{Los Angeles}
  \country{USA}}

\author{Tian Ye}
\authornotemark[1]
\orcid{0000-0001-9298-8906}
\email{	tye69227@usc.edu}
\affiliation{%
  \institution{University of Southern California}
  \city{Los Angeles}
  \country{USA}}

\author{Cauligi Raghavendra}
\orcid{0009-0001-8966-7738}
\email{raghu@usc.edu}
\affiliation{%
  \institution{University of Southern California}
  \city{Los Angeles}
  \country{USA}}

\author{Viktor Prasanna}
\orcid{0000-0002-1609-8589}
\email{prasanna@usc.edu}
\affiliation{%
  \institution{University of Southern California}
  \city{Los Angeles}
  \country{USA}}

\renewcommand{\shortauthors}{Wickramasinghe et al.}

\newcommand{\calX}{{\mathcal{X}}}
\newcommand{\calY}{{\mathcal{Y}}}
\newcommand{\calH}{{\mathcal{H}}}
\newcommand{\calC}{{\mathcal{C}}}
\newcommand{\calA}{{\mathcal{A}}}
\newcommand{\calB}{{\mathcal{B}}}
\newcommand{\calF}{{\mathcal{F}}}
\newcommand{\calR}{{\mathcal{R}}}
\newcommand{\calZ}{{\mathcal{Z}}}
\newcommand{\calM}{{\mathcal{M}}}
\newcommand{\VC}{{\text{\rm VCdim}}}
\newcommand{\blue}[1]{{\color{blue}#1}}
\newcommand{\R}{\mathbb{R}}
\newcommand{\domain}{\text{domain}}
\newcommand{\E}{\mathbb{E}}
\newcommand{\KL}{D_\text{KL}}
\newcommand{\CE}{\ell_\text{ce}}
\newcommand{\cbr}[1]{\left\{#1\right\}}
\newcommand{\sbr}[1]{\left\[#1\right\]}
\newcommand{\rbr}[1]{\left(#1\right)}
\newcommand{\inner}[2]{\left\langle #1,#2 \right\rangle}
\newcommand{\order}{\ensuremath{\mathcal{O}}}
\newcommand{\sign}{{\text{\rm sign}}}
\newcommand{\indicator}{\mathbf{1}}
\newcommand{\statoracle}{{\text{\rm STAT}}}
\newcommand{\exampleoracle}{{\text{\rm EX}}}
\newcommand{\sqdim}{{\text{\rm SQ-DIM}}}
\newcommand{\ldim}{{\text{\rm Ldim}}}
\newcommand{\twonorm}[1]{\left\| #1\right\|_2}
\newcommand{\onenorm}[1]{\left\| #1\right\|_1}
\newcommand{\norm}[1]{\left\| #1\right\|}
\newcommand{\abs}[1]{\left| #1\right|}

\begin{abstract}
Convolutional Neural Networks (CNNs) have achieved state-of-the-art accuracy in Synthetic Aperture Radar (SAR) Automatic Target Recognition (ATR). However, their high computational cost, latency, and memory footprint make its deployment challenging on resource-constrained platforms such as small satellites. 
While adversarial robustness is critical for real-world SAR ATR, it is often overlooked in system-level optimizations. Achieving both robustness and inference efficiency requires a unified framework that considers adversarially trained models together with hardware constraints. 
In this paper, we present a model-hardware co-design framework for CNN-based SAR ATR that integrates robustness-preserving model compression with FPGA accelerator design. 
The compression stage includes hardware-guided structured pruning, where a hardware performance model derived from the FPGA design predicts the pruning impact on latency and resource usage. This enables the generation of Pareto-optimal models that improve hardware efficiency under user-defined objectives, while maintaining adversarial robustness within a predefined tolerance.
We design an FPGA accelerator with channel-aware Processing Element (PE) allocation that supports both fully pipelined streaming and temporal resource-reuse architectures. An automated design generation flow efficiently maps the compressed models to optimized FPGA implementations. Experiments on the widely used MSTAR and FUSAR-Ship datasets across three CNN architectures show that our framework produces models up to 18.3$\times$ smaller with 3.1$\times$ fewer MACs while preserving robustness. Our FPGA implementation achieves up to 68.1$\times$ (6.4$\times$) lower inference latency and up to  169.7$\times$ (33.2$\times$) better energy efficiency compared to CPU (GPU) baselines, demonstrating the effectiveness of the proposed co-design framework for robust and efficient SAR ATR on FPGA platforms.
\end{abstract}




\maketitle
\section{Introduction}
Synthetic Aperture Radar (SAR) is a key technique in remote sensing and defense applications due to its ability to capture imagery regardless of lighting or weather conditions. Automatic Target Recognition (ATR) based on SAR imagery is widely used in real-world systems such as military missions and autonomous decision platforms, where reliable and real-time decision-making is required. 
Unlike optical image recognition, SAR ATR relies on radar signals that are sensitive to speckle noise, phase errors, and intentional interference. These characteristics make SAR ATR especially vulnerable to adversarial perturbations. Robustness against such perturbations is therefore indispensable for real deployment.

Deep learning models, especially Convolutional Neural Networks (CNNs)~\cite{cnn1,cnn2,cnn3,cnn4}, have significantly improved SAR ATR accuracy. However, two main challenges remain for practical deployment: \textbf{adversarial robustness} and \textbf{inference efficiency} on hardware platforms.
First, SAR ATR is highly vulnerable to adversarial perturbations~\cite{fgsm,pgd,sar-attack1,sar-attack2,sar-attack3,sar-attack4,sar-attack5, ye2023adversarial, ye2023realistic}.
At inference time, an attacker can manipulate a model's prediction by adding carefully crafted perturbations to the input SAR image. These perturbations can be imperceptible to human observers, yet effective in causing SAR ATR models to misclassify the target. Such vulnerability is unacceptable in safety-critical scenarios and motivates adversarial robustness of SAR ATR models.
Second, efficient inference under hardware resource constraints is essential. We define \textit{inference efficiency} in terms of inference latency, number of multiply-accumulate operations (MACs), model size, and energy efficiency. 
SAR ATR has two common deployment scenarios: on-board SAR image processing on edge platforms~\cite{small_sat1, small_sat4, small_sat5} and SAR image processing at data centers~\cite{nasa_supercompute}. In both scenarios, low inference latency and high energy efficiency are critical.
Adversarial training has been shown effective to improve robustness against adversarial perturbations~\cite{pgd}, but typically increases model complexity and computational cost. At the same time, efficient deployment on FPGA platforms requires careful consideration of architectural constraints such as dataflow, parallelism, and on-chip resource allocation.
Prior works on SAR model compression or FPGA acceleration have largely focused on improving inference efficiency without explicitly considering adversarial robustness~\cite{wang2021boosting, chen2018slim,babu2024optimizing, gnn, wickramasinghe2025cnn-att4}. As a result, adversarially trained models often remain too expensive for practical deployment. 
Furthermore, model pruning strategies guided by FLOPs or parameter counts do not necessarily translate into proportional reductions in latency or FPGA resource usage, since actual inference efficiency depends on architectural factors such as dataflow and resource mapping.


To address these issues, we propose a model-hardware co-design framework for efficient deployment of adversarially trained SAR ATR models on FPGA. 
The framework integrates adversarial training, hardware-guided structured pruning, and parameterized accelerator design within a unified workflow. In this framework, we focus on preserving the robustness of models while improving hardware efficiency through model compression and hardware-aware design. 
Specifically, we introduce (1) a hardware-guided structured pruning algorithm that balances robustness preservation and hardware-specific efficiency, (2) an FPGA accelerator design combined with an analytical hardware performance model to enable rapid evaluation during pruning, (3) a model-hardware coupling methodology in which hardware constraints inform pruning decisions and pruning results configure the accelerator generation through an automated design flow, and (4) parameterized modular compute engines that support both fully pipelined streaming and temporal resource-reuse architectures, enabling scalability across FPGA platforms with varying resource budgets and deployment of deeper CNN models under strict hardware constraints.

Given a user-specified CNN architecture, a SAR dataset, and target FPGA metadata such as available DSPs and BRAMs, our framework first applies adversarial training to obtain a robust baseline model. It then performs hardware-guided structured pruning, where pruning decisions are informed by both saliency scores and an analytical hardware performance model derived from our accelerator design. The performance model estimates the cost of each channel in terms of MACs, latency, and FPGA resources, enabling the pruning algorithm to explore robustness-efficiency trade-offs under user-specified objectives and constraints. The pruning process produces a set of Pareto-optimal candidate models, models where no other candidate improves hardware efficiency without reducing robustness, from which the user can select based on their application requirements. The selected model is further quantized and implemented on FPGA using parameterized high-level synthesis (HLS) templates. We evaluate our framework on the widely used MSTAR~\cite{mstar} and FUSAR-Ship~\cite{hou2020fusar} datasets to demonstrate generality across different SAR domains.
Overall, the framework provides an end-to-end co-design flow from robust and efficient model generation to optimized accelerator implementation. 
To the best of our knowledge, this is the first model-hardware co-design framework that systematically integrates adversarially trained SAR ATR models with model compression and FPGA accelerator generation.
Our main contributions are:
\begin{itemize}
    \item We propose a model-hardware co-design framework for CNN-based SAR ATR that improves inference efficiency on FPGA for adversarially trained models while preserving adversarial robustness.
    \item We develop a robustness-aware hardware-guided structured pruning algorithm that couples robustness saliency with hardware cost metrics. 
    By balancing saliency scores and per-channel hardware cost estimates, the pruning algorithm explores robustness-efficiency trade-offs under a predefined robustness tolerance.
    \item We design an accelerator architecture composed of modular compute engines with channel-aware PE allocation, which can be instantiated as a fully pipelined streaming dataflow accelerator for large FPGAs or as a temporal resource‑reuse accelerator for smaller FPGAs, enabling portability and efficient inference of compressed CNNs.
    \item We implement an automated design generation flow using parameterized high-level synthesis (HLS) templates, which seamlessly maps the compressed models to optimized FPGA implementations.
    \item Our framework generates models up to 18.3$\times$ smaller with 3.1$\times$ fewer MACs, and FPGA implementations achieving up to 68.1$\times$ (6.4$\times$) lower inference latency and 169.7$\times$ (33.2$\times$) better energy efficiency over CPU (GPU) baselines, while maintaining adversarial robustness.
\end{itemize}

\section{Background}
\label{sec:background}
\subsection{Adversarial  Robustness}


While deep learning models have shown remarkable performance in SAR ATR, recent studies~\cite{fgsm,pgd,sar-attack1,sar-attack2,sar-attack3,sar-attack4,sar-attack5} have revealed their vulnerability to adversarial attacks. These attacks involve adding perturbations to input images at inference time and causing the model to produce incorrect predictions. 
The general principle behind adversarial attacks is to find input perturbations that maximizes the model's prediction loss. Let $f_\theta$ denote a SAR image classifier with trained parameters $\theta$, $x$ be an input image, $y$ be its ground truth label, and $L$ be the prediction loss (e.g., cross-entropy). An adversarial example $\tilde{x}$ is defined as
\[
    \tilde{x}=\arg\max_{\tilde{x}\in \calB(x,\varepsilon)} L(f_\theta(\tilde{x}), y),
\]
where $\calB(x,\varepsilon)=\{x'\mid\norm{x'-x}\le\varepsilon\}$ constrains perturbation magnitude. A widely used method to compute $\tilde{x}$ is Projected Gradient Descent (PGD)~\cite{pgd}, which iteratively updates
\[
    \tilde{x}^{(t+1)}=\Pi_{\calB(x,\varepsilon)}[\tilde{x}^{(t)}+\eta\cdot\text{sgn}(\nabla_xL(f_\theta(\tilde{x}^{(t)}),y))],
\] where $\tilde{x}^{(0)}=x$, $\eta$ is the step size, and $\Pi$ is the projection operator that enforces the perturbation constraint.
In this paper, we define \textbf{adversarial robustness} as the classification accuracy of a model against PGD attacks under a specified perturbation budget, which is a widely used robustness metric.
This vulnerability is especially critical for SAR ATR, because SAR imagery is inherently noisy and of a low contrast, making classifiers more sensitive to perturbations than in optical imagery. Moreover, SAR ATR is often deployed in safety-critical applications such as defense missions and autonomous decision systems, where adversarial attacks can cause severe consequences. Therefore, adversarial robustness is more than essential for SAR ATR.

Among defense strategies against adversarial attacks, adversarial training~\cite{trades} has proven particularly effective. 
It formulates the learning problem as: 
\[
    \min_\theta \E_{(x,y)\in D}\max_{\tilde{x}\in\calB(x,\epsilon)}L(f_\theta(\tilde{x}),y),
\]
where $D$ is the training dataset. This enhances robustness by generating adversarial examples on-the-fly during training and using them to update the model, encouraging the learning of stable and reliable features that are less sensitive to perturbations. 
Due to the complicated learning objective, adversarial training often requires wider or deeper networks to improve the accuracy-robustness trade-off, thereby increasing inference cost.
Most prior work~\cite{wei2024moar,xu2021adversarial,sar-attack1} on adversarial robustness of SAR ATR focuses on model architectures or training algorithms, with little attention to system-level efficiency. This reveals a critical gap: to achieve high robustness and low inference latency on resource-constrained platforms, it requires the joint optimization of the robust model and the hardware accelerator. To address this gap, we propose a framework that integrates adversarial training and model optimization with model-hardware co-design, delivering robust CNN-based SAR ATR while meeting resource budgets and providing low inference latency.

\subsection{CNN-based SAR ATR}

CNN architectures remain the dominant and most widely used approach for SAR ATR~\cite{sar-atr-survey}, making them a natural choice for hardware acceleration frameworks. While alternative approaches like Vision Transformers (ViTs) and Graph Neural Networks (GNNs) have shown promise in recent research, they face significant challenges in SAR ATR, including high computational complexity, reliance on large labeled datasets for effective training, and limited generalizability~\cite{fein2023benchmarking, wickramasinghe2025cnn-att4}. In contrast, CNN-based models offer proven effectiveness and are well established in real-world SAR ATR systems.
CNNs for SAR ATR have evolved from standard networks to more specialized designs~\cite{sar-atr-survey}.
In the early years of exploring deep learning methods for SAR ATR, standard CNNs pre-trained on optical image datasets, such as AlexNet~\cite{alexnet}, VGGNet~\cite{vgg}, and ResNet~\cite{resnet}, were adopted due to their demonstrated success in mainstream computer vision applications~\cite{cnn-performance-analysis}. These models were typically fine-tuned on SAR datasets like MSTAR~\cite{mstar}, yielding competitive classification accuracy and establishing the viability of CNN-based approaches for SAR target recognition~\cite{Fu2018cnn2, Soldin2018cnn3, anas2020cnn4}.
This foundation led to the development of customized CNN architectures specifically designed to address SAR-specific challenges, including limited labeled data, noise robustness, and unique imaging characteristics~\cite{aconvnet, qnet, sar-net, zhao2018cnn8, huang2020lightweight, mf-sarnet, umbrellanet, m-net, wickramasinghe2025cnn-att4}. 

\section{Model-Hardware Co-design Framework}
\begin{figure}[ht]
\centerline{\includegraphics[width=\textwidth]{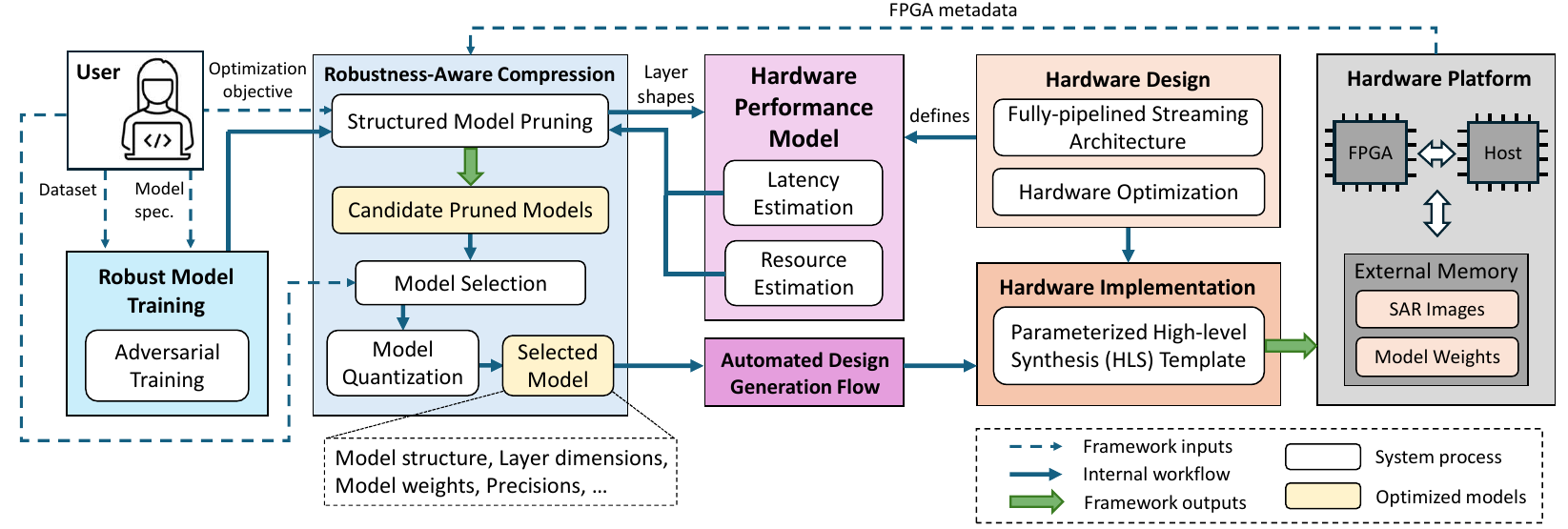}}
\caption{Overview of the proposed framework.}
\Description{}
\label{fig:overview}
\end{figure}

Figure~\ref{fig:overview} presents an overview of the proposed framework, which provides a complete pipeline for developing adversarially robust and hardware-efficient SAR ATR models for FPGA platforms. 
The framework takes four user-specified inputs: (1) a CNN model architecture defined in PyTorch, (2) a SAR image dataset (e.g., MSTAR~\cite{mstar}), (3) an optimization objective related to hardware (e.g., minimizing MACs, latency, or FPGA resource usage), and (4) target FPGA metadata (e.g., available DSPs, BRAMs, etc.). Based on these inputs, the framework automatically trains, prunes, and quantizes a set of robust model candidates, evaluates their robustness and inference efficiency, and outputs: (1) a set of FPGA-feasible pruned and quantized models spanning various robustness-efficiency trade-offs, enabling the user to select the one that best fits their deployment requirements, and (2) an HLS-based hardware implementation enabling FPGA acceleration of model inference.

At the core of our co-design framework are two key features that enable joint optimization of model and hardware. First, a hardware performance model, derived from our FPGA accelerator design, provides accurate latency and resource estimates that guide the model pruning process. Second, an automated design generation flow facilitates seamless adaptation of the accelerator to implement the resulting compressed model architecture. Together, these two components tightly couple model optimization with hardware implementation, ensuring coordinated model-hardware co-design rather than independent, sequential optimization.
Overall, the framework consists of four key components: robust model training and compression, hardware performance model, hardware design, and hardware implementation. Each part progressively refines the model for robust and efficient deployment on hardware.

\textbf{Robust Model Training and Compression: }
The user-defined CNN model and SAR image dataset are first processed by our PyTorch front-end to train a robust classifier using adversarial training with PGD~\cite{pgd}. This results in a baseline robust model with FP32 precision that serves as the input to the model compression. The robust model is then compressed through three steps. (1) \textit{Structured Model Pruning}: It generates a set of model candidates with different robustness-efficiency trade-offs under the user-defined optimization objective and hardware constraints. An analytical hardware performance model is used to facilitate the model pruning.
(2) \textit{Model Selection}: Users are provided with estimations on each candidate's robustness and inference efficiency, and are allowed to select a model that meets their application requirements. (3) \textit{Model Quantization}: It applies mixed-precision quantization to reduce the model size and inference cost. In particular, we cast the weights of convolutional and fully connected layers to INT8, while other layers remain in FP32 to preserve accuracy.

\textbf{Hardware Performance Model: }
To enable hardware-guided model optimization, we develop a hardware performance model (Section \ref{sec:perf_model}). Given layer dimensions (e.g., input/output channels, kernel size, feature map size), this model outputs \textit{hardware cost estimates}, including per-channel MACs, inference latency, and FPGA resource usage (DSPs and BRAMs). These estimates can be queried efficiently without synthesis, closely matching measured results.
During pruning, the performance model is repeatedly queried to estimate the cost reduction from removing each channel, enabling hardware-guided model compression. After pruning, the same model is used again to evaluate the overall cost of all candidate models meeting the target FPGA's resource constraints, so that the users can select a preferred model to configure the HLS templates. 

\textbf{Hardware Design:} 
To support efficient inference of robust and compressed SAR ATR models, the framework includes an FPGA accelerator design based on parameterized HLS templates (Section~\ref{sec:hw_design}). The architecture supports both fully pipelined streaming dataflow and temporal resource-reuse implementations to accommodate different FPGA resource budgets. It comprises modular compute engines for convolution, max-pooling, and General Matrix Multiplication (GEMM), and supports mixed-precision execution using INT8 arithmetic with 32-bit accumulation and FP32 operations where required. The accelerator employs a channel-aware PE allocation strategy in which the number of instantiated PEs is configured according to the pruned channel dimensions, subject to a design-specific maximum.
When channel dimensions exceed this limit, channel folding is applied to maintain efficient hardware utilization. The framework explores candidate PE limits during pruning to align model structure with hardware resource constraints.

\textbf{Hardware Implementation: }
Once a pruned and quantized model is selected, the framework employs an \textit{automated design generation flow} (Section \ref{sec:adf}) and parameterized HLS templates to instantiate the corresponding accelerator design. It compiles the design with the AMD Vitis~\cite{vitis} toolchain to generate an FPGA bitstream. Finally, the framework loads the model weights along with SAR image inputs onto the FPGA for low-latency and robust inference. Our co-designed solution enables real-time and adversarially robust SAR ATR inference under hardware constraints, benefiting from the tight integration between model optimization and automated hardware customization.

\section{Robustness-Aware Model Compression}
\label{sec:model-compression}

\subsection{Adversarial Training Setup}
Our framework begins by training a robust SAR ATR model using adversarial training. The user provides a SAR dataset and a configuration specifying the model architecture $f$. We apply adversarial training based on PGD as described in Section~\ref{sec:background}, where adversarial examples are generated on-the-fly during the training. Specifically, we minimize 
\(
\frac{1}{n}\sum_{i=1}^n\CE\left(f_\theta(\tilde{x}_i),y_i\right),  
\)
where $\theta$ denotes trainable model parameters, $(x_i,y_i)$ is the $i$-th clean sample and label, and $\tilde{x}_i$ is the adversarial example generated by PGD with perturbation budget $\varepsilon=8/255$ under $\ell_\infty$-norm (assuming pixel values of the SAR image $x_i$ are normalized to $[0,1]$). This perturbation budget is widely adopted in adversarial robustness research~\cite{pgd} as it balances imperceptibility to humans while being strong enough to reliably test model robustness. This procedure trains the model to be robust to adversarial perturbations at inference time. The resulting adversarially trained model, referred to as the \textit{initial robust model}, serves as the starting point for the subsequent model compression stages.

\subsection{Hardware-Guided Structured Pruning}

Starting from the initial robust model, we perform structured pruning to improve inference efficiency on FPGA while preserving adversarial robustness throughout the pruning process. 
Our pruning algorithm features two aspects: (1) \textbf{Structured model pruning for easy FPGA mapping}, which removes entire output channels from convolutional and fully connected layers. This structured pruning reduces computation and memory footprint while preserving regularity in data access and computation patterns, ensuring that pruned models can be naturally mapped to FPGA accelerators. (2) \textbf{Hardware guidance}, where an analytical hardware performance model derived from our accelerator design (Section~\ref{sec:hardware}) guides each pruning step. For every remaining channel, the model predicts the reduction in the hardware cost with respect to a user-specified objective (e.g., latency, or DSP/BRAM usage) if that channel were removed. We combine this predicted cost reduction with a saliency score computed on the adversarially trained model to prioritize channels for pruning and account for both robustness preservation and inference efficiency. The model robustness is explicitly evaluated after each pruning step, and the pruning proceeds only while the robustness degradation remains within a tolerance.

\begin{figure}[h]
\centerline{\includegraphics[width=0.7\linewidth]{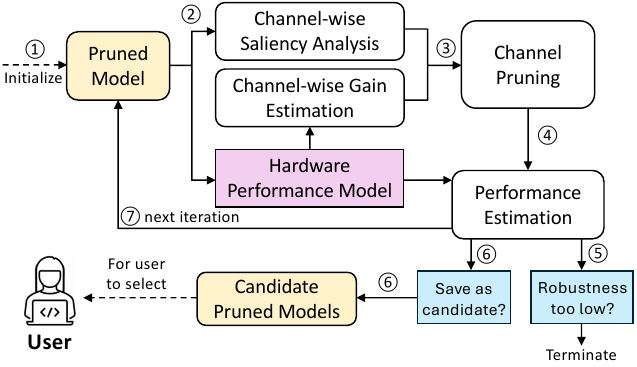}}
\caption{Overview of hardware-guided structured pruning.}
\Description{}
\label{fig:pruning}
\end{figure}

\begin{algorithm}[t]
\small
\caption{Hardware-Guided Structured Pruning}
\label{alg:prune}
\KwIn{
  Initial robust model $f_\theta$, Robustness drop threshold $\tau$,\\
   Objective $\mathcal{C}$, Saliency function $S$, Performance model $\mathcal{H}$,\\   
  Checkpoint decay factor $\rho$}
\KwOut{Candidate model set $\mathcal{M}_\text{cand}$}

$R_\text{base} \leftarrow$ PGD$(f_\theta)$ \tcp*{Initial robustness}
$O_\text{base} \leftarrow \mathcal{H}(f_\theta, \mathcal{C})$ \tcp*{Initial hardware cost}
$O_\text{next} \leftarrow \rho \cdot O_\text{base}$\;
$\mathcal{M}_\text{cand} \leftarrow \{(f_\theta,R_\text{base},O_\text{base})\}$\;

\While{True}{
  \ForEach{remaining channel $(l,c)$}{
  $g_{l,c} \leftarrow \Delta\mathcal{H}(f_\theta, (l,c), \mathcal{C})$ \tcp*{Predicted cost reduction if $(l,c)$ is removed}
    $S_{l,c} \leftarrow S(f_\theta,(l,c))$ \tcp*{Saliency}
      $P_{l,c} \leftarrow g_{l,c} / (S_{l,c}+\epsilon)$ \tcp*{Priority score}
  }
  $(l^*,c^*) \leftarrow \arg\max P_{l,c}$\;
  Prune channel $(l^*,c^*)$ from $f_\theta$\;
  $R_\text{cur} \leftarrow$ PGD$(f_\theta)$\;
  $O_\text{cur} \leftarrow \mathcal{H}(f_\theta, \mathcal{C})$\;
  \If{$R_\text{base}-R_\text{cur} > \tau R_\text{base}$}{
        break\;
  }
  \If{$O_\text{cur} \le O_\text{next}$}{
        Append $(f_\theta,R_\text{cur},O_\text{cur})$ to $\mathcal{M}_\text{cand}$\;
        $O_\text{next} \leftarrow \rho \cdot O_\text{cur}$\;
  }
}
\Return{$\calM_\text{cand}$}
\end{algorithm}

Figure~\ref{fig:pruning} illustrates the pruning workflow, which begins with the initial robust model (Step 1). 
In each iteration, we (i) compute a channel-wise saliency on the adversarially trained model, and (ii) estimate the hardware cost reduction (referred to as \textit{hardware gain}) using the analytical performance model (Step 2). 
These analyses determine the next channel to prune (Steps 3--4). After each step of pruning, we re-evaluate the model's robustness and hardware cost (Steps 4--5). If the robustness remains within the tolerance and sufficient progress is made towards the hardware objective, we record the current model as a candidate checkpoint (Step 6). The loop continues until the robustness degradation exceeds the tolerance, yielding a set of candidate models that expose the robustness-efficiency trade-off for deployment. 

\textbf{Channel-Wise Saliency Analysis: }
To preserve robustness during pruning, we estimate the importance of each output channel using saliency scores. 
A key point is that the saliency is computed on an adversarially trained model that resists attacks under the perturbation budget. As a result, the measured saliency score acts as a practical proxy that discourages the removal of channels that the initial robust model significantly relies on. 
For the $l$-th convolutional layer and its $c$-th output channel, we compute a saliency score $S_{l,c}$ using one of the following definitions:
\begin{enumerate}[leftmargin=*]
    \item[(1)] $\ell_p$-norm of weights:
    \(
        S_{l,c}^\text{weight}=\norm{w_{l,c}}_p
    \),
    where $p=1$ or $p=2$, and $w_{l,c}$ is the weights for that channel. The intuition is that the channel with lower magnitudes in its associated weights $w_{l,c}$ will have less impact on the output of the model. 
    \item[(2)] Activation mean:
    \(
        S_{l,c}^\text{act} = \E_{x\in D}[\text{mean}(|z_{l,c}(x))|]
    \),
    where $z_{l,c}(x)$ represents the activation of the $c$-th output channel of the $l$-th convolutional layer for a given input SAR image $x$. The intuition is that if a channel's output is closer to zero, it will be less important.
    \item[(3)] First-order Taylor saliency: 
    \[
        S_{l,c}^\text{taylor}=\abs{\E_{(x,y)\in D_\text{batch}}\left[\frac{\partial L(f_\theta(x),y)}{\partial z_{l,c}}\cdot z_{l,c}\right]},
    \]
    where the expectation is taken over a batch of samples. This score estimates the immediate effect on the loss $L$ if the channel were removed. 
\end{enumerate}

\textbf{Channel-Wise Gain Estimation: }
To prune the model for better inference efficiency on FPGA, our framework allows the user to specify an optimization objective $\mathcal{C}$, such as MACs, inference latency, or FPGA resource usage (e.g., DSP, BRAM). 
For each channel $(l,c)$, we estimate a \textit{hardware gain} $g_{l,c}$ as the predicted reduction in the chosen hardware cost $\mathcal{C}$ when that channel is removed. In other words, a larger $g_{l,c}$ indicates a greater efficiency improvement.
If the objective is MACs, the gain can be computed analytically as
\(
    g_{l,c}^{\text{mac}} = C_{l-1} \times K_l^2 \times H_l^{\text{out}} \times W_l^{\text{out}},
\)
where $C_{l-1}$ is the number of input channels, $K_l$ is the kernel size, and $H_l^{\text{out}}$ and $W_l^{\text{out}}$ are the height and width of the output feature map. For latency and resource usage (DSP/BRAM), we rely on the analytical hardware performance model derived from our FPGA design (Section~\ref{sec:hardware}) to provide fast estimates without requiring synthesis.

\textbf{Model Pruning: }
To balance efficiency and robustness, we iteratively prune the channel with the highest priority score, defined as: 
\(
P_{l,c}=g_{l,c}/(S_{l,c}+\epsilon), 
\)
where $g_{l,c}$ is the predicted hardware gain, $S_{l,c}$ is the robustness saliency score, and $\epsilon$ is a small constant for numerical stability. 
This implements a trade-off between hardware efficiency and robustness: channels offering 
larger predicted hardware cost reduction are prioritized to be removed only when their saliency is low, which discourages the removal of channels that the model robustness strongly relies on. After each step of pruning, we explicitly re-evaluate robustness under PGD to ensure that the saliency of remaining channels is aligned with actual robustness preservation. The user specifies the optimization objective $\mathcal{C}$ but does not need to set a fixed target value for it. This is important in practice because the user may lack prior knowledge of how much pruning the model can tolerate without severe robustness degradation.
Instead, the pruning process maintains a candidate set $\calM_\text{cand}$ of intermediate models that will be fine-tuned, quantized, and evaluated later. The pruning process terminates once robustness degradation exceeds the tolerance $\tau$, producing Pareto-optimal candidate models in terms of robustness and the optimization objective, from which the users can select a model that best fits their requirements.

\textbf{Exponential Checkpointing:}
To decide when to record a candidate model, we use an exponential checkpointing strategy. A pruned model is saved only if the hardware cost drops by a factor $\rho$ compared to the most recently saved candidate in $\calM_\text{cand}$. Let $O_\text{base}$ denote the hardware cost (e.g., latency) of the initial robust model. A copy of the model is saved when the current cost $O_\text{cur}$, estimated by the performance model, satisfies $O_\text{cur}\le \rho^k\cdot O_\text{base}$, $k=1,2,...$. This approach can reduce redundancy and capture key points along the robustness-efficiency trade-off curve.

\textbf{Stopping Criterion: }
To determine when to terminate the pruning, we evaluate the pruned model's robustness against PGD-20 on a small evaluation dataset after each pruning step. Let $R_\text{base}$ be the robustness of the initial robust model and $\Delta_\text{max}=\tau\cdot R_\text{base}$ be the maximum allowed robustness degradation (we use $\tau=0.05$). If the current robustness $R_\text{cur}$ satisfies $R_\text{base}-R_\text{cur}>\Delta_\text{max}$, the pruning algorithm is stopped.

Overall, our pruning algorithm enables the framework to produce efficient models with bounded robustness degradation, and allows the user to explore robustness-efficiency trade-offs.
Our approach also differs from prior works~\cite{sehwag2020hydra,yu2019autoslim}, which either require fixed sparsity targets or optimize with expensive search procedures. We instead offer an efficient approach that integrates structured pruning with robustness and hardware-guided optimization.

\subsection{Fine-Tuning and Quantization}

After structured pruning, each candidate model is fine-tuned to recover accuracy and robustness. This involves a few epochs of adversarial training using a reduced learning rate. 
We then apply post-training quantization to further reduce the size of each candidate model and enable more efficient hardware acceleration. Specifically, we convert the precision of all convolutional and fully connected layers from FP32 to INT8 with commonly used symmetric per-tensor quantization for weights and asymmetric per-layer quantization for activations~\cite{quant}. 
We adopt post-training quantization to maintain compatibility with FPGA toolchains, and to keep the model compression pipeline lightweight without introducing additional training complexity (e.g., quantization-aware adversarial training).

\subsection{Evaluation and Pareto Frontier Selection}

After fine-tuning and quantization, each candidate model is evaluated on a validation set. We assess natural accuracy and adversarial robustness against PGD-20 attack under $\ell_\infty$-norm with $\varepsilon=8/255$. Besides, we also use the same hardware performance model (Section~\ref{sec:perf_model}) to estimate their expected latency and resource usage. Users then select from a set of Pareto-optimal candidates with respect to robustness and inference efficiency, and the chosen model is passed to our parameterized HLS templates for FPGA implementation.

\section{Hardware Design and Implementation}
\label{sec:hardware}
\subsection{Architecture Design and Optimizations}
\label{sec:hw_design}

\begin{figure}[h]
\centerline{\includegraphics[width=\linewidth]{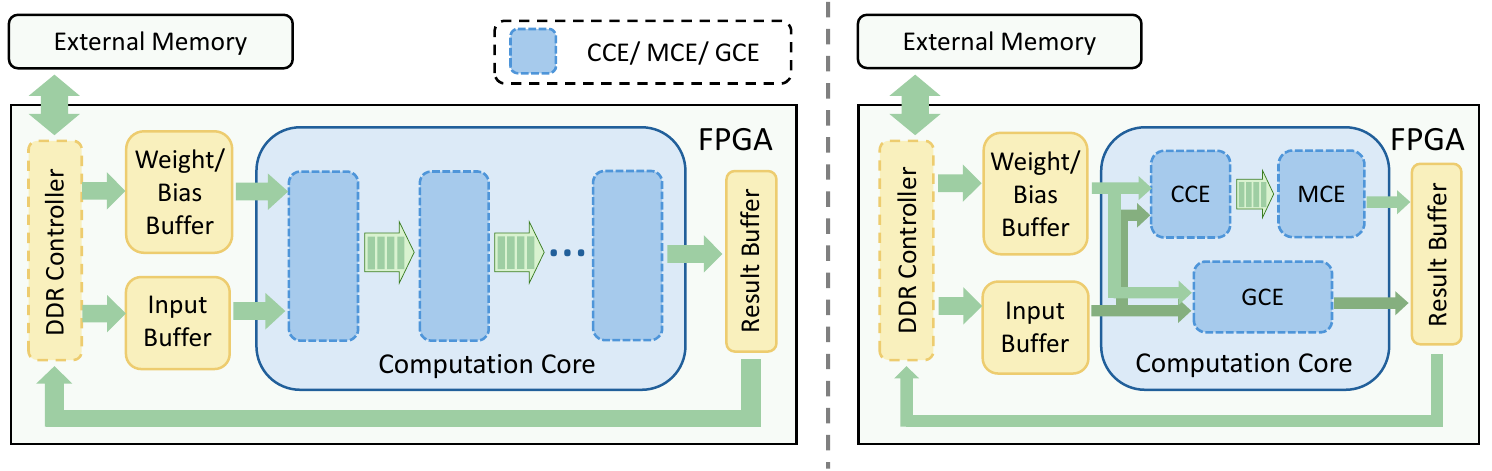}}
\caption{Overview of the hardware architecture. (Left:  Streaming dataflow design, Right: Temporal resource-reuse design)}
\Description{}
\label{fig:hw_overview}
\end{figure}

\subsubsection{Overview of the Hardware Architecture}
The hardware system consists of a host device and an FPGA. The host transfers input SAR image data to the FPGA’s on-chip memory, while model weights and biases are loaded onto on-chip buffers. The main compute units are implemented using parameterized HLS templates that can be instantiated in two modes depending on the available resources.
For large FPGAs, we use the templates to instantiate a fully pipelined streaming dataflow accelerator. As shown in Figure~\ref{fig:hw_overview}, the architecture consists of a sequence of compute engines connected by FIFO streams, forming a pipelined chain that forwards intermediate results directly between layers. This streaming design enables continuous data flow and avoids off-chip accesses. Once inference is complete, the results are written back to the FPGA's external memory.
For resource-constrained FPGAs, we use the templates to instantiate a temporal, resource‑reuse accelerator (Figure~\ref{fig:hw_overview}). In this mode, compute engines are reused across layers, and all intermediate results are written back to off-chip memory. Our parameterized template design preserves the same model mapping flow while enabling deployment across devices with different resource budgets. The modularity of the computation core also allows flexible implementation of different layers based on the model architecture. 




\subsubsection{Parameterized Compute Engine Design}
The computation core consists of three main compute engines, Convolution Compute Engine (CCE), Max-Pooling Compute Engine (MCE), and GEMM Compute Engine (GCE), implemented using parameterized HLS templates. Depending on the selected model, these templates can be configured with 
different layer settings, such as kernel size, number of input and output channels, and feature map dimensions, and are then instantiated by our framework. In streaming mode, compute engines are connected through FIFO streams to enable inter-layer pipelining without intermediate off-chip memory accesses. In temporal resource-reuse mode, a single instance of each compute engine is time-multiplexed across layers, and intermediate feature maps are written to and read from external memory between layer executions. For convolutional layers followed by max-pooling, the CCE streams its outputs directly to the MCE through on-chip FIFOs before results are written to external memory. If a pooling layer is not present, the CCE output bypasses the MCE stage.



\begin{figure}[ht]
\centerline{\includegraphics[width=\linewidth]{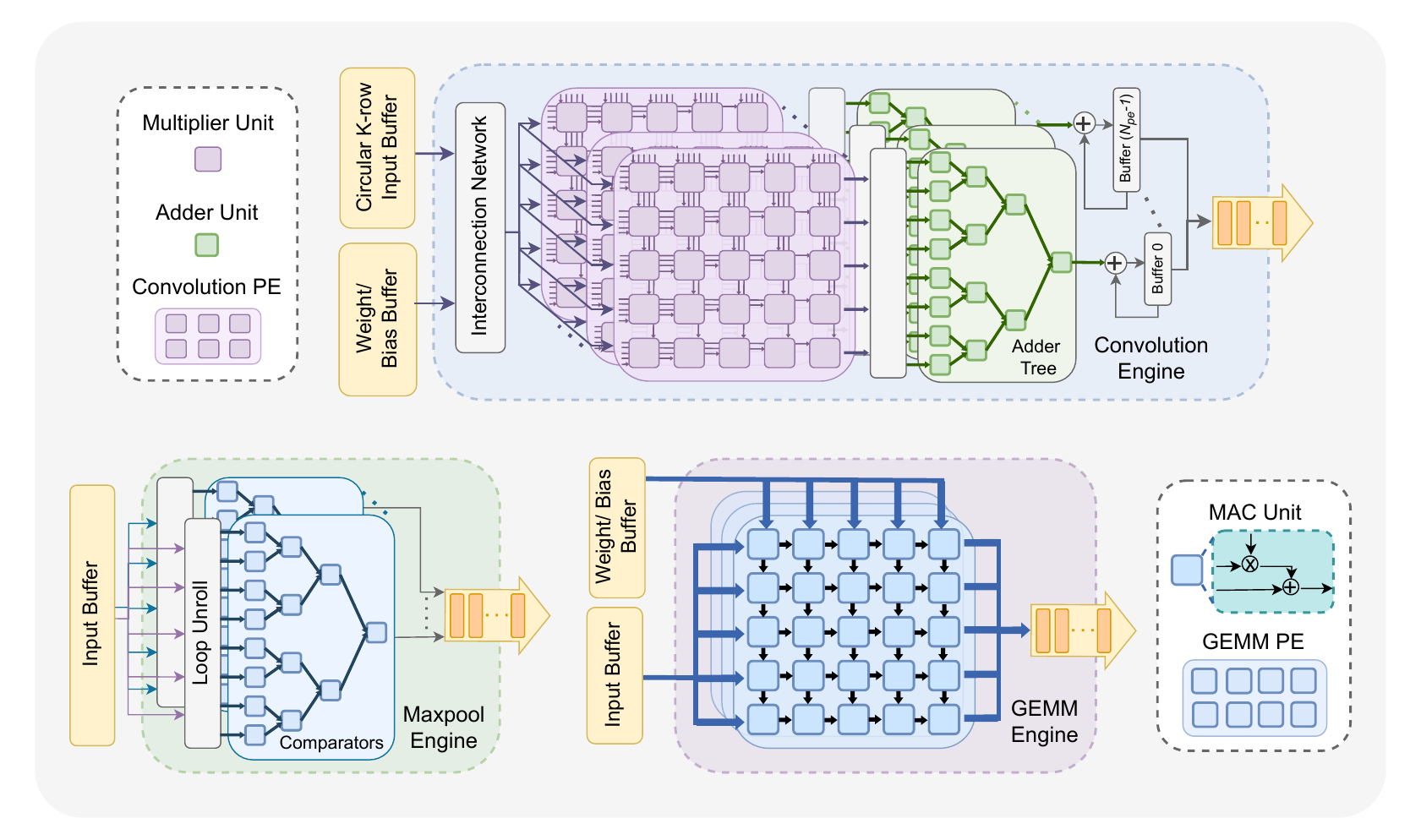}}
\caption{Convolution, max-pooling and GEMM compute engines.}
\Description{}
\label{fig:conv}
\end{figure}

\textit{\textbf{Convolution Compute Engine (CCE):}} Figure \ref{fig:conv} depicts the architecture of the CCE. The CCE instantiates $N_\text{pe}$ parallel convolution processing elements (PE), each supporting $K \times K$ multipliers for the convolution kernel size $K$. 
The number of PEs, $N_\text{pe}$, is typically configured to match the number of output channels ($C_l$) for a given convolution layer. Depending on the resource availability, we apply channel folding by limiting $N_\text{pe} = N_\text{pe}^\text{max} < C_l$ and processing channels in multiple passes. 
An input repacking stage reorders channel groups between folds to match the layout required by the next convolution layer.
The CCE reads activations through a $K$-row circular line buffer mechanism with a rotating head pointer that selectively overwrites exactly \textit{stride} rows corresponding to the new data entering the next sliding window.
The CCE uses loop unrolling and array partitioning to maximize parallelism enabling $N_\text{pe} \times K \times K $ parallel multiplications. Each PE maps its $K \times K$ multipliers to DSP blocks, while intermediate feature maps and weights are buffered using on-chip BRAMs with array partitioning to sustain parallel access.
The outputs from the multipliers are fed into $N_\text{pe}$ adder trees, each producing a partial sum for a specific output channel. Once the convolution results for all $N_\text{pe}$ output channels are computed, they are forwarded to the MCE. 



\textit{\textbf{Max-Pooling Compute Engine (MCE):}} The MCE operates at the same parallel granularity as the CCE, consuming $N_\text{pe}$ parallel outputs per cycle. In streaming mode, max-pooled outputs are forwarded directly to the next compute engine, whereas in temporal reuse mode, they are written back to external memory. The MCE maximizes throughput by computing max-pooling over streamed input feature maps. As shown in Figure~\ref{fig:conv}, it operates in a fully-pipelined manner, overlapping input buffering, max-pool window reduction, and requantization to achieve a high processing rate of pixels.
The MCE comprises $N_\text{pe}$ parallel comparator trees, each supporting a $K_\text{m} \times K_\text{m}$ window, where $K_\text{m}$ denotes the max-pooling window size. Each comparator tree reduces its inputs to a single maximum value, yielding $N_\text{pe}$ outputs. These outputs are then requantized using precomputed scales and zero-points and streamed to the next layer. 


\textit{\textbf{GEMM Compute Engine (GCE):}} In streaming mode, the GCE consumes streamed activations from preceding layers and produces outputs without intermediate buffering to off-chip memory. 
In temporal reuse mode, matrix operands are loaded from external memory into on-chip buffers prior to each GEMM invocation.
The GCE accelerates matrix multiplications found in linear layers and similar operators. It uses a systolic array where MAC units are connected in a grid for pipelined matrix multiplication. 
The outer loops are pipelined to initiate one row-column pair multiplication per cycle. 
Multiple GEMM PEs can be instantiated in parallel for independent matrix multiplications. 
The systolic array dimensions are parameterized according to layer dimensions and available resources.

\subsubsection{Channel-Aware PE Allocation Strategy}
To maintain improved hardware utilization under structured pruning, we establish a direct mapping between model channel dimensions and PEs. For a convolutional layer with $C_l$ output channels, the number of instantiated PEs, $N_\text{pe}$, is configured at compile time according to the pruned model structure. When sufficient resources are available, a one-to-one mapping is used ($N_\text{pe} = C_l$), ensuring that each output channel is computed by a dedicated PE and eliminating the inefficiencies of fixed PE architectures where pruned channels leave hardware underutilized. When the available resources cannot support full parallelism, channel folding is applied by constraining $N_\text{pe}$ to a predefined design-specific maximum, $N_\text{pe}^\text{max}$ ($N_\text{pe}^\text{max} \in \{8,16,32,64\}$ and $N_\text{pe}^\text{max} < C_l$). In the temporal design, a single instance of the CCE is synthesized with $N_\text{pe} = N_\text{pe}^{\max}$ fixed across all layers. In cases where $N_\text{pe} < C_l$, output channels are partitioned into $\lceil C_l / N_\text{pe} \rceil$ folds, and each fold is processed sequentially using the same set of PEs across both convolution and max-pooling stages. For channel counts that are not exact multiples of $N_\text{pe}$, the final fold may contain fewer active channels. This results in reduced PE utilization only during the last fold of the corresponding layer. Since latency is typically dominated by layers with larger channel counts, the overall throughput degradation due to channel imbalance remains limited. To ensure correct dataflow between layers under folding, an input repacking stage reorganizes channel groups so that the streamed data layout matches the expected input ordering of the subsequent convolution layer. This mechanism preserves compatibility across varying channel dimensions generated by pruning and channel folding.


\subsubsection{System-Level Hardware Optimizations}

The hardware design integrates several optimizations for efficiency and adaptability. 
The modular compute engines adapt to pruned architectures through template-based instantiation, enabling efficient inference of models with varying configurations. For the selected candidate model, the framework generates a model-specific hardware instance at synthesis time using parameterized HLS templates.
The design supports quantized convolution and max pooling by incorporating requantization logic for intermediate activations. Convolutions use 8-bit activations and weights with 32-bit accumulation to preserve numerical stability. We apply inline requantization after max pooling to keep the latency of the convolution unaffected. Batch normalization is fused into the convolution operation, eliminating the need for separate computations. We use a $K$-row circular line buffer with a rotating head pointer to eliminate redundant data movement during stride operations.
The design employs fine-grained pipelining within each engine. In streaming mode, compute engines are connected through inter-layer FIFO streams. In temporal reuse mode, intermediate feature maps are written to external memory between layer executions except when a max-pooling layer immediately follows a convolution.
To support high-throughput data movement, the FIFO streams are implemented using custom vector data types, allowing multiple outputs, such as $N_\text{pe}$ parallel results, to be written simultaneously. Parallelism is further enhanced by partitioning the data buffers and applying loop unrolling.

\subsection{Hardware Performance Model}
\label{sec:perf_model}
We develop an analytical performance model for our accelerator architecture to predict latency and resource usage (validated in Section~\ref{sec:perf_model_val}). We formulate equations based on our compute engine designs with empirically calibrated constants.
Although the constants are calibrated for the target FPGA, the modeling methodology is architecture-driven rather than platform-specific. The equations are derived from pipeline structure, loop initiation intervals, and channel folding behavior. When targeting a different FPGA family, device-dependent constants can be recalibrated while preserving the same analytical framework.
Since convolution and max-pooling layers dominate runtime and are most affected by pruning, we model the latency (in clock cycles) and resource usage of CCEs and MCEs. The design maps multiply-accumulate operations to dedicated DSP units and uses BRAMs for input and activation buffering, making DSPs and BRAMs the primary resource constraints in our estimation. 

\subsubsection{Latency Estimation} 

The CCE follows a pipelined architecture and is composed of two primary stages: initial input buffer loading and convolution computation with circular buffer reads. Its cycle-level latency can be modeled as: \(t_\text{conv} = t_\text{input\_load} + t_\text{compute}\)
\begin{gather*}
t_\text{input\_load} = K_l \times W_l^\text{in} \times II_\text{input} + D_\text{input} \\
t_\text{compute} = \lceil\frac{C_l}{N_\text{pe}}\rceil \left[ H_l^\text{out} \times W_l^\text{out}\times(t_\text{loop} + t_\text{ov}) + (H_l^\text{out}-1) \times t_\text{buffer}\right] \\
t_\text{loop} = C_{l-1}\times II_\text{conv}+D_\text{conv}, \quad \quad \quad t_\text{buffer} = S_{l}\times W_l^\text{in}\times II_\text{b}+D_\text{b}
\end{gather*}
where $II$ and $D$ denote the initiation interval and the pipeline depth, respectively. Based on our hardware design, we set $II_\text{input}= II_\text{conv} = II_\text{b} = 1$, $D_\text{input}=D_\text{b}=3$, and $D_\text{conv}=7$. An overhead $t_\text{ov}=7$ accounts for any memory access delays or loop control logic. $K_l$ and $S_l$ are the kernel size and stride. For the first convolution layer (with a single-channel SAR input), we partition along the input width dimension eliminating the $W_l^\text{in}$ factor from input loading.
The cycle-level latency of MCE can be modeled as:
\[
t_\text{maxpool} = \lceil\frac{C_l}{N_\text{pe}}\rceil \times (H_l^\text{in}+2P) \times (W_l^\text{out}+2P)\times II_\text{maxpool} + D_\text{maxpool}
\]
where $P$ is the padding size for max-pooling. We set $II_\text{maxpool}=6$ to allow limited BRAM banking and empirically set $D_\text{maxpool}=50$. In both CCE and MCE, we scale the total number of cycles by a factor of $\frac{C_l}{N_\text{pe}}$, reflecting the degree of channel-wise parallelism.


\subsubsection{Resource Requirement Estimation (DSP/BRAM)} 
We estimate the DSP and BRAM usage for each CCE as:
\begin{gather*}
DSP_\text{conv} =N_\text{pe} \times K_l \times K_l/\rho_1, \quad \quad \quad
BRAM_\text{conv} = C_{l-1} \times K_l
\end{gather*}
The factor ${\rho}_1=1.56$ is an empirically determined constant, indicating the DSP packing factor. The BRAM usage depends on the array partitioning strategy applied to the input. 
The DSP and BRAM usage for each MCE can be modeled as:
\begin{gather*}
DSP_\text{maxpool} = N_\text{pe} /{\rho}_2 + d_\text{ov},\quad \quad \quad BRAM_\text{maxpool} = N_\text{pe}
\end{gather*}
where ${\rho}_2 = 1.6$ is the empirically determined DSP packing factor and $d_\text{ov} = 4$ is a fixed DSP overhead for other shared operations. Although max-pooling itself does not use DSPs, the MCE employs DSPs for requantizations and other logic. 

\begin{figure}[ht]
\centerline{\includegraphics[width=\linewidth]{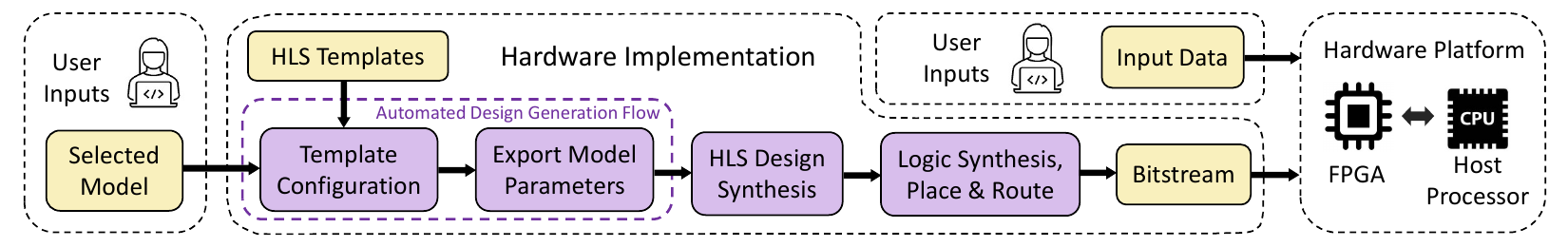}}
\caption{Overview of the hardware implementation workflow.}
\Description{}
\label{fig:hw_deploy}
\end{figure}

\subsection{Hardware Implementation}
\label{sec:adf}

\begin{Code}
\begin{lstlisting}[
  caption  = {Core of parameterized HLS template for CCE.},
  label    = {lst:hls-core},
  language = C++,
  basicstyle=\ttfamily\scriptsize,
  backgroundcolor=\color{backcolour},
  keywordstyle=\color{keywordblue},
  commentstyle=\color{codegreen},
  morekeywords={[2]pragma},
  keywordstyle={[2]\color{pragmagreen}},
  literate={##}{{{\color{pragmagreen}\#}}}1,
  frame=none,
  numbers=none]
template<int IH, int IW, int OH, int OW,
         int IC, int OC, int K, int S, int P, int PE>
void cnn_layer(hls::stream<vec_in>& in_s, 
               hls::stream<vec_out>& out_s,
               const wt_t W[OC][IC][K][K], 
               const bias_t B[OC]) {
  /* preload: W  -> Wbuf[FOLD][PE][IC][K][K]
              B  -> Bbuf[FOLD][PE]
     stage input -> InBuf[IC][K][IW] */
  const int FOLD = (OC + PE - 1) / PE;
  for (int f = 0; f < FOLD; ++f) 
    for (int oh = 0; oh < OH; ++oh) 
      for (int ow = 0; ow < OW; ++ow) {
        acc_t sum[PE];  // complete array partition
        for (int pe = 0; pe < PE; ++pe)  // unrolled
          sum[pe] = Bbuf[f][pe];
        for (int c = 0; c < IC; ++c) {  // pipelined II=1
          for (int kh = 0; kh < K; ++kh) {  // unrolled  
            for (int kw = 0; kw < K; ++kw) {  // unrolled 
              for (int pe = 0; pe < PE; ++pe) {  // unrolled
                sum[pe] += Wbuf[f][pe][c][kh][kw] *
                           InBuf[c][(oh*S+kh)%K][ow*S-P+kw];
        }}}}
        write_out(sum);   // stream PE results
      }
      for (int s = 0; s < S; ++s) {  // buffer update
        int h = compute_head_pointer();
        for (int iw = 0; iw < IW; ++iw) { // pipelined II=1
           for (int ic = 0; ic < IC; ++ic) { // unrolled
              InBuf[ic][h][iw] = read_in(in_s); }}}
}
\end{lstlisting}
\end{Code}

The final stage of our framework implements the hardware accelerator design for FPGA deployment. The HLS templates define parameterized compute engines that are instantiated at synthesis time according to the selected pruned model. Listing \ref{lst:hls-core} presents the core of the parameterized HLS template for the CCE. The input and output feature map dimensions, channels, and convolution parameters such as kernel size, stride, and padding, are defined as configurable template parameters. Weights and biases are preloaded into local buffers, while the input feature maps are fed to the compute engine through a streaming interface. Fine-grained pipelining is applied to both the input loading and the convolution compute loop. To enable concurrent computation of $N_\text{pe} \times K \times K $ multiply-accumulate operations, we apply loop unrolling and array partitioning to the innermost loops. All outputs produced simultaneously by the PEs are packed into a vector data structure and written to the output stream in a single cycle. 

\textbf{Automated Design Generation Flow: } As shown in Figure~\ref{fig:hw_deploy}, once a Pareto-optimal pruned model is selected, our lightweight automated design generation flow maps the compressed model specification into synthesizable FPGA hardware through a systematic two-stage process. The first stage focuses on updating the configurations of the modular HLS templates. Given the per-layer channel dimensions for the selected pruned model as inputs, the generation flow automatically derives the corresponding template parameters and updates the design. Parameters unaffected by pruning
remain unchanged from the original model specification. The second stage handles weight and quantization parameter extraction. The flow extracts all model weights, biases, quantization scales, and zero points from the compressed model, then transforms this data to align with the architecture's memory access patterns. This includes converting data types to match the mixed-precision execution model and organizing arrays to support the parallel processing elements in each compute engine. Following design generation, the resulting synthesizable C/C++ design is passed to the Vitis HLS toolchain to generate RTL, perform synthesis and place-and-route, and produce a bitstream, which is loaded onto the FPGA together with the input data and model parameters.

\section{Experiments}

\subsection{Experimental Setup}
\paragraph{Datasets} 
Publicly available SAR datasets with labeled targets are limited due to high cost of data acquisition and data sensitivity. Consequently, we select two representative and widely used SAR datasets that capture common SAR ATR scenarios involving ground vehicles and maritime targets. Specifically, we evaluate our framework on two SAR datasets: the widely used Moving and Stationary Target Acquisition and Recognition (MSTAR) dataset~\cite{mstar} for ground vehicle recognition, and the FUSAR-Ship dataset~\cite{hou2020fusar} for maritime target recognition. For MSTAR, we use the standard 10-class benchmark with 2,747 training and 2,425 test images. For FUSAR-ship, we select a 5-class subset of ship chips with 500 training images and 4,006 test images. All images are $128\times 128$ single-channel intensity maps of SAR returns, and pixel values are normalized to $[0,1]$.



\textit{Model Architectures.}
Our experiments focus on CNN architectures, which remain the most widely used and dominant modeling paradigm for SAR ATR. Compared to transformer-based models, CNNs are better suited to the limited size of available SAR datasets and exhibit stronger inductive biases for learning local scattering patterns.
To demonstrate the generality of our approach, our experiments use three representative CNN architectures for SAR ATR: (1) \textit{Attn-CNN}~\cite{wickramasinghe2025cnn-att4}, a lightweight model enhanced with attention mechanisms, which reflects the current state-of-the-art among CNN-based SAR ATR, (2) \textit{AlexNet}~\cite{alexnet}, a widely adopted baseline in SAR ATR research that serves as a conventional architecture of earlier CNNs, and (3) \textit{Two-Stream}~\cite{huang2020lightweight}, a model designed for SAR ATR that extracts multilevel features through parallel local and global convolution streams demonstrating superior recognition accuracy. All models are implemented in PyTorch and adversarially trained to serve as the initial robust model prior to compression.

\textit{Model Training and Attack Settings.} 
We adopt Projected Gradient Descent (PGD) under the $\ell_\infty$ threat model for adversarial training and evaluation. The perturbation budget is set to $\varepsilon=8/255$, a commonly used setting in adversarial robustness research. During training, adversarial examples are generated with 10 steps of PGD with a step size of $2/255$. After structured pruning, candidate models are fine-tuned for 10 epochs using the same adversarial training procedure. For robustness evaluation, we employ a stronger 20-step PGD attack with the same perturbation budget and step size. This setting follows standard white-box evaluation protocol and allows us to consistently measure robustness preservation before and after compression.


\textit{Hardware Platforms. }
We implement our hardware design using parameterized High-Level Synthesis (HLS) templates on an AMD Alveo U280 FPGA,  synthesized with the AMD Vitis 2022.2 toolchain~\cite{vitis}.
Our work jointly optimizes adversarial robustness and hardware efficiency for SAR ATR, a research direction unexplored in existing works. To our knowledge, no directly comparable co-design frameworks exist. Therefore, we evaluate our FPGA co-design against CPU and GPU baseline implementations.
Specifically, we measure inference performance on a state-of-the-art CPU server with an AMD EPYC 9754 processor and on an NVIDIA RTX 6000 Ada GPU. On CPU, inference is performed in FP32 and INT8 using PyTorch~\cite{pytorch} with \texttt{fbgemm} backend~\cite{fbgemm}. On GPU, inference is performed in FP32 using PyTorch and in INT8 via TensorRT 10.13~\cite{tensorrt}, with energy measured by NVML~\cite{nvml}. For fair comparison, identical model architectures, pruning configurations, and quantization precisions are used across all platforms.
All hardware specifications are summarized in Table~\ref{tab:specifications}.


\begin{table}[ht]
\centering
\caption{Specifications of hardware platforms}
\begin{adjustbox}{max width=0.8\textwidth}
\begin{tabular}{cccc}
\toprule
\textbf{Platform} & \begin{tabular}[|c|]{@{}c@{}} \textbf{CPU}\\ AMD EPYC 9754 \end{tabular} & \begin{tabular}[|c|]{@{}c@{}} \textbf{GPU} \\ NVIDIA RTX 6000 Ada \end{tabular} & \begin{tabular}[|c|]{@{}c@{}} \textbf{FPGA} \\ AMD Alveo U280 \end{tabular}  \\
\midrule
Technology & TSMC 5 nm & TSMC 4 nm & TSMC 16 nm \\ \midrule
Frequency & 2.25 GHz & 2.505 GHz & 300 MHz \\ \midrule
Peak Performance & 4.61 TFLOPS & 91.1 TFLOPS & 0.96 TFLOPS\\ \midrule
On-chip Memory & 256 MB L3 cache & 96 MB & 45 MB\\ \bottomrule
\end{tabular}
\end{adjustbox}
\label{tab:specifications}
\end{table}

\textit{Performance Metrics.}
For a comprehensive evaluation of the trade-offs between robustness and inference efficiency achieved by our framework, we evaluate multiple metrics. To quantify \textit{robustness}, we report the classification \textbf{accuracy} on clean inputs, and \textbf{adversarial robustness} against a 20-step PGD attack with $\ell_\infty$ perturbations bounded by $\varepsilon=8/255$. To quantify \textit{inference efficiency}, we measure: (1) computational complexity through the number of multiply-accumulate operations (\textbf{MACs}) per inference, (2) \textbf{model size} as the product of parameter count and quantization bit precision, (3) \textbf{inference latency} defined as the time required to process a single SAR image, and (4) \textbf{energy efficiency} defined as the consumed energy per image. We measure single-image inference latency with a batch size of 1, as it best reflects SAR ATR's operational setting~\cite{bhanu1986automatic}. Additionally, we also report the \textbf{DSP and BRAM} utilization on the target hardware for evaluating robustness-efficiency trade-offs. FPGA inference latency and resource usage are obtained using the AMD Vitis Analyzer~\cite{vitis_analyzer}, while power measurements are reported using the AMD Power Design Manager~\cite{pdm}.

\begin{table}[ht]
\centering
\caption{Inference performance comparison on CPU, GPU, and FPGA.}
\begin{adjustbox}{max width=\textwidth}\begin{tabular}{c|c|cc|cc|cc}
\Xhline{1pt}
& & & & \multicolumn{2}{c|}{\textbf{MSTAR}} & \multicolumn{2}{c}{\textbf{FUSAR-Ship}} \\
\cline{5-6}\cline{7-8}
\textbf{Model} & \textbf{Device} & \textbf{Pruned} & \textbf{Quantized} &
\makecell{\textbf{Latency} \textbf{(ms)}} &
\makecell{\textbf{Energy Efficiency}  \textbf{(mJ/Image)}} &
\makecell{\textbf{Latency} \textbf{(ms)}} &
\makecell{\textbf{Energy Efficiency} \textbf{(mJ/Image)}} \\
\Xhline{1pt}
\multirow{10}{*}{\textbf{Attn-CNN}} 

& \multirow{4}{*}{CPU} 
& \ding{55} & \ding{55} & 10.177 (23.24$\times$) & 791 (68.2$\times$) & 13.963 (24.28$\times$) & 1081 (70.9$\times$) \\
&& \ding{51} & \ding{55} & 7.634 (17.43$\times$) & 594 (51.1$\times$) & 10.713 (18.63$\times$) & 829 (54.4$\times$) \\
&& \ding{55} & \ding{51} & 5.259 (12.01$\times$) & 409 (35.2$\times$) & 5.934 (10.32$\times$) & 460 (30.2$\times$) \\
&& \cellcolor{lightblue}\ding{51} & \cellcolor{lightblue}\ding{51} & \cellcolor{lightblue}4.361 (9.96$\times$)  & \cellcolor{lightblue}339 (29.2$\times$) & \cellcolor{lightblue}5.969 (10.38$\times$) & \cellcolor{lightblue}463 (30.4$\times$)\\
\cline{2-8}

& \multirow{4}{*}{GPU} 
& \ding{55}   & \ding{55} & 2.822 (6.44$\times$) & 197 (17.0$\times$) & 2.782 (4.84$\times$) & 214 (14.0$\times$)\\
&& \ding{51}  & \ding{55} & 2.454 (5.60$\times$) & 172 (14.8$\times$) & 2.466 (4.29$\times$) & 185 (12.1$\times$)\\
&& \ding{55}   & \ding{51} & 0.525 (1.20$\times$) & 25 (2.1$\times$) & 0.583 (1.01$\times$) & 33 (2.2$\times$)\\
&& \cellcolor{lightblue}\ding{51}   & \cellcolor{lightblue}\ding{51} & \cellcolor{lightblue}0.492 (1.12$\times$)  & \cellcolor{lightblue}28 (2.3$\times$) & \cellcolor{lightblue}0.559 (0.97$\times$) & \cellcolor{lightblue}31 (2.1$\times$)\\
\cline{2-8}

& \multirow{2}{*}{FPGA}
& \ding{55}  & \ding{51}  & 1.080 (2.47$\times$) & 29 (2.5$\times$) & 1.080 (1.88$\times$) & 29 (1.9$\times$)\\
&& \cellcolor{lightgr}\ding{51}  & \cellcolor{lightgr}\ding{51}  & \cellcolor{lightgr}0.438 (1.00$\times$) & \cellcolor{lightgr}12 (1.0$\times$) & \cellcolor{lightgr}0.575 (1.00$\times$) & \cellcolor{lightgr}15 (1.0$\times$)\\

\Xhline{1pt}
\multirow{10}{*}{\textbf{AlexNet}}  

& \multirow{4}{*}{CPU} 
& \ding{55}   & \ding{55}   & 10.903 (49.79$\times$)  & 682 (122.2$\times$) & 13.339 (68.06$\times$) & 848 (169.7$\times$) \\
&& \ding{51}  & \ding{55}   & 4.521 (20.64$\times$)  & 283 (50.7$\times$) & 5.268 (26.88$\times$) & 407 (81.4$\times$)\\
&& \ding{55}  & \ding{51} & 1.360 (6.21$\times$)   & 85 (15.2$\times$) & 1.795 (9.16$\times$) & 139 (27.8$\times$)\\
&& \cellcolor{lightblue}\ding{51}  & \cellcolor{lightblue}\ding{51}  & \cellcolor{lightblue}1.267 (5.79$\times$) & \cellcolor{lightblue}79 (14.2$\times$) & \cellcolor{lightblue}1.658 (8.46$\times$) & \cellcolor{lightblue}128 (25.6$\times$)\\
\cline{2-8}

& \multirow{4}{*}{GPU} 
& \ding{55}   & \ding{55}& 1.078 (4.92$\times$) & 170 (30.4$\times$) & 1.069 (5.45$\times$) & 166 (33.2$\times$)\\
&& \ding{51}  & \ding{55} & 0.981 (4.48$\times$) & 154 (27.6$\times$) & 1.010 (5.15$\times$) & 95 (19.0$\times$)\\
&& \ding{55}   & \ding{51}& 0.405 (1.85$\times$)  & 29 (4.8$\times$) & 0.467 (2.38$\times$) & 30 (6.0$\times$)\\
&& \cellcolor{lightblue}\ding{51}   & \cellcolor{lightblue}\ding{51}& \cellcolor{lightblue}0.395 (1.80$\times$)  & \cellcolor{lightblue}32 (5.3$\times$) & \cellcolor{lightblue}0.467 (2.38$\times$) & \cellcolor{lightblue}23 (4.6$\times$)\\
\cline{2-8}

& \multirow{2}{*}{FPGA} 
& \ding{55}    & \ding{51}    & 0.984 (4.49$\times$) & 25 (4.5$\times$) & 0.984 (5.02$\times$) & 25 (5.0$\times$)\\
&& \cellcolor{lightgr}\ding{51}    & \cellcolor{lightgr}\ding{51}    & \cellcolor{lightgr}0.219 (1.00$\times$) & \cellcolor{lightgr}6 (1.0$\times$) & \cellcolor{lightgr}0.196 (1.00$\times$) & \cellcolor{lightgr}5 (1.0$\times$)\\

\Xhline{1pt}
\multirow{10}{*}{\textbf{Two-Stream}}  

& \multirow{4}{*}{CPU} 
& \ding{55}   & \ding{55}   &  3.117 (10.75$\times$)  & 240 (33.1$\times$) & 5.288 (19.37$\times$) & 410 (60.1$\times$)\\
&& \ding{51}  & \ding{55}   & 2.286 (7.88$\times$)  & 176 (24.3$\times$) & 4.256 (15.59$\times$) & 329 (48.2$\times$)\\
&& \ding{55}  & \ding{51} &  1.243 (4.29$\times$)  & 96 (13.2$\times$) & 1.544 (5.66$\times$) & 120 (17.6$\times$)\\
&& \cellcolor{lightblue}\ding{51}  & \cellcolor{lightblue}\ding{51}  & \cellcolor{lightblue}1.165 (4.02$\times$) & \cellcolor{lightblue}90 (12.4$\times$) & \cellcolor{lightblue}1.294 (4.74$\times$)& \cellcolor{lightblue}100 (14.7$\times$)\\
\cline{2-8}

& \multirow{4}{*}{GPU} 
& \ding{55}   & \ding{55}& 0.945 (3.26$\times$)  & 68 (9.4$\times$) & 1.033 (3.78$\times$) & 77 (11.3$\times$)\\
&& \ding{51}  & \ding{55} & 0.972 (3.35$\times$) & 70 (9.7$\times$) & 0.847 (3.10$\times$) & 63 (9.2$\times$)\\
&& \ding{55}   & \ding{51}& 0.379 (1.31$\times$)  & 20 (2.9$\times$) & 0.406 (1.49$\times$) & 25 (3.6$\times$)\\
&& \cellcolor{lightblue}\ding{51}   & \cellcolor{lightblue}\ding{51}& \cellcolor{lightblue}0.374 (1.29$\times$)  & \cellcolor{lightblue}16 (2.3$\times$) & \cellcolor{lightblue}0.416 (1.52$\times$)& \cellcolor{lightblue}21 (3.0$\times$)\\

\cline{2-8}

& \multirow{2}{*}{FPGA}
& \ding{55}    & \ding{51}    & 0.426 (1.47$\times$) & 11 (1.5$\times$) & 0.426 (1.56$\times$) & 11 (1.6$\times$)\\
&& \cellcolor{lightgr}\ding{51}    & \cellcolor{lightgr}\ding{51}    & \cellcolor{lightgr}0.290 (1.00$\times$) & \cellcolor{lightgr}7 (1.0$\times$) & \cellcolor{lightgr}0.273 (1.00$\times$) & \cellcolor{lightgr}7 (1.0$\times$)\\

\Xhline{1pt}
\end{tabular}
\end{adjustbox}
\label{tab:performance}
\end{table}

\subsection{Inference Efficiency Evaluation}

\textbf{\textit{Latency, Energy Efficiency:}} We evaluate the inference performance of robust SAR ATR models generated by our framework on FPGA and compare them with CPU and GPU implementations. Table~\ref{tab:performance} reports results on MSTAR and FUSAR-Ship across three CNN architectures. 
Overall, the FPGA consistently delivers substantial acceleration. For AlexNet, it achieves 49.8--68.1$\times$ speedup over CPU and 4.9--5.5$\times$ over GPU than unpruned and unquantized baselines. Compared to pruned and quantized versions on CPU and GPU, it remains up to 8.5$\times$ and 2.4$\times$ faster. Similar trends are observed on Two-Stream. For Attn-CNN, the FPGA significantly outperforms CPU, while speedups over GPU are modest (0.97--1.12$\times$), likely due to limited quantized datapath optimizations.
We also observe that pruning has limited effect on GPU latency, as the pruned channel dimensions result in inefficient kernel utilization, and existing GPU libraries cannot efficiently map irregular pruned channel sizes. In contrast, pruning strongly benefits FPGA inference, as our accelerator adapts its compute units to the pruned model structure. 
Quantized-only CPU/GPU implementations sometimes outperform pruned and quantized ones for the same reason, whereas the FPGA fully exploits both pruning and quantization. 
Furthermore, the FPGA provides up to 169.7$\times$ and 33.2$\times$ better energy efficiency than CPU and GPU, respectively, a critical advantage for resource-constrained platforms.
Together, these results validate that our co-design framework effectively produces robust and efficient SAR ATR across different datasets and CNNs.

\textbf{\textit{MACs, Model Size:}} Table~\ref{tab:ablation} presents the accuracy, robustness, MACs, and model size of the three CNN models across MSTAR and FUSAR-Ship datasets. The results demonstrate that pruning alone significantly reduces both MACs and model size. Applying quantization to the pruned models further decreases model size with only a marginal drop in robustness, confirming that pruning and quantization together contribute to model efficiency while preserving robustness. For example, AlexNet's pruned and quantized model achieves an 18.3$\times$ reduction in size and 3.1$\times$ reduction in MACs, compared to the unpruned and unquantized baseline, while incurring only a 2.3\% drop in robustness. Similar trends are observed for Attn-CNN and Two-Stream models across both datasets, where the framework produces models with substantially reduced MACs and model size while preserving the robustness.

\begin{table}
\centering
\caption{Evaluation of accuracy, robustness, MACs, and model size. (A: Accuracy, R: Robustness)}
{
\begin{adjustbox}{max width=\textwidth}
\begin{tabular}{c|cc|cccc|cccc}
\Xhline{0.8pt}
& & & \multicolumn{4}{c|}{\textbf{MSTAR}} & \multicolumn{4}{c}{\textbf{FUSAR-Ship}} \\
\cline{4-11}
 \textbf{Model} & \textbf{Pruned} & \textbf{Quantized} & \textbf{A (\%)} 
 & \textbf{R (\%)} 
 & \textbf{MACs ($\times10^8$)} & \textbf{Model Size (MB)} & \textbf{A (\%)} 
 & \textbf{R (\%)} 
 & \textbf{MACs ($\times10^8$)} & \textbf{Model Size (MB)} \\ 

\Xhline{0.8pt}
\multirow{4}{*}{\textbf{Attn-CNN}} 
& \ding{55} & \ding{55} & 97.57 & 75.71 & $5.85$ (FP32) & 1.96 ($1.00\times$) & 54.52 & 42.36 & $5.85$ (FP32) & 1.96 ($1.00\times$) \\ 
& \ding{55} & \ding{51} & 97.03 & 75.33 & $5.83$ (INT8) $+ 0.01$ (FP32) & 0.72 ($\downarrow2.72\times$)  & 54.47 & 42.66 & $5.83$ (INT8) $+ 0.01$ (FP32) & 0.72 ($\downarrow2.72\times$) \\ 
& \ding{51} & \ding{55} & 97.57 & 74.06 & $1.23$ (FP32) & 1.24 ($\downarrow1.58\times$) & 54.22 & 39.82 & $3.27$ (FP32) & 1.71 ($\downarrow1.15\times$) \\
& \ding{51} & \ding{51} & 97.48 & 73.77 & $1.22$ (INT8) $+ 0.01$ (FP32) & 0.54 ($\downarrow3.63\times$) & 52.75 & 40.74 & $3.26$ (INT8) $+ 0.01$ (FP32) & 0.66 ($\downarrow2.97\times$)\\
\hline
\multirow{4}{*}{\textbf{AlexNet}} 
& \ding{55} & \ding{55} & 92.29 & 75.42 & $2.36$ (FP32) & 228.12 ($1.00\times$) & 44.96 & 39.04 & $2.36$ (FP32) & 228.12 ($1.00\times$) \\ 
& \ding{55} & \ding{51} & 92.08 & 74.72 & $2.36$ (INT8) & 57.03 ($\downarrow4.00\times$) & 44.63 & 39.02 & $2.36$ (INT8) & 57.03 ($\downarrow4.00\times$) \\ 
& \ding{51} & \ding{55} & 92.49 & 73.94 & $0.77$ (FP32) & 50.00 ($\downarrow4.56\times$) & 45.03 & 38.92 & $1.77$ (FP32) & 71.51 ($\downarrow3.19\times$)\\
& \ding{51} & \ding{51} & 92.41 & 73.11 & $0.77$ (INT8) & 12.50 ($\downarrow18.25\times$) & 44.56 & 38.99 & $1.77$ (INT8) & 17.88 ($\downarrow12.76\times$)\\
\hline
\multirow{4}{*}{\textbf{Two-Stream}} 
& \ding{55} & \ding{55} & 94.76 & 83.79 & $2.36$ (FP32) & 1.01 ($1.00\times$) & 52.05 & 41.81 & $2.36$ (FP32) & 1.01 ($1.00\times$) \\ 
& \ding{55} & \ding{51} & 94.68 & 83.34 & $2.36$ (INT8) & 0.25 ($\downarrow4.00\times$) & 52.22 & 41.79 & $2.36$ (INT8) & 0.25 ($\downarrow4.00\times$) \\ 
& \ding{51} & \ding{55} & 94.68 & 83.34 & $0.62$ (FP32) & 0.72 ($\downarrow1.39\times$) & 51.32 & 41.16 & $0.75$ (FP32) & 0.96 ($\downarrow1.05\times$) \\
& \ding{51} & \ding{51} & 94.06 & 82.47 & $0.62$ (INT8) & 0.18 ($\downarrow5.56\times$) & 51.40 & 41.29 & $0.75$ (INT8) & 0.24 ($\downarrow4.20\times$) \\
\Xhline{0.8pt}
\end{tabular}
\end{adjustbox}
}
\label{tab:ablation}
\end{table}

\begin{figure}[htbp]
\centering
\includegraphics[width=\textwidth]{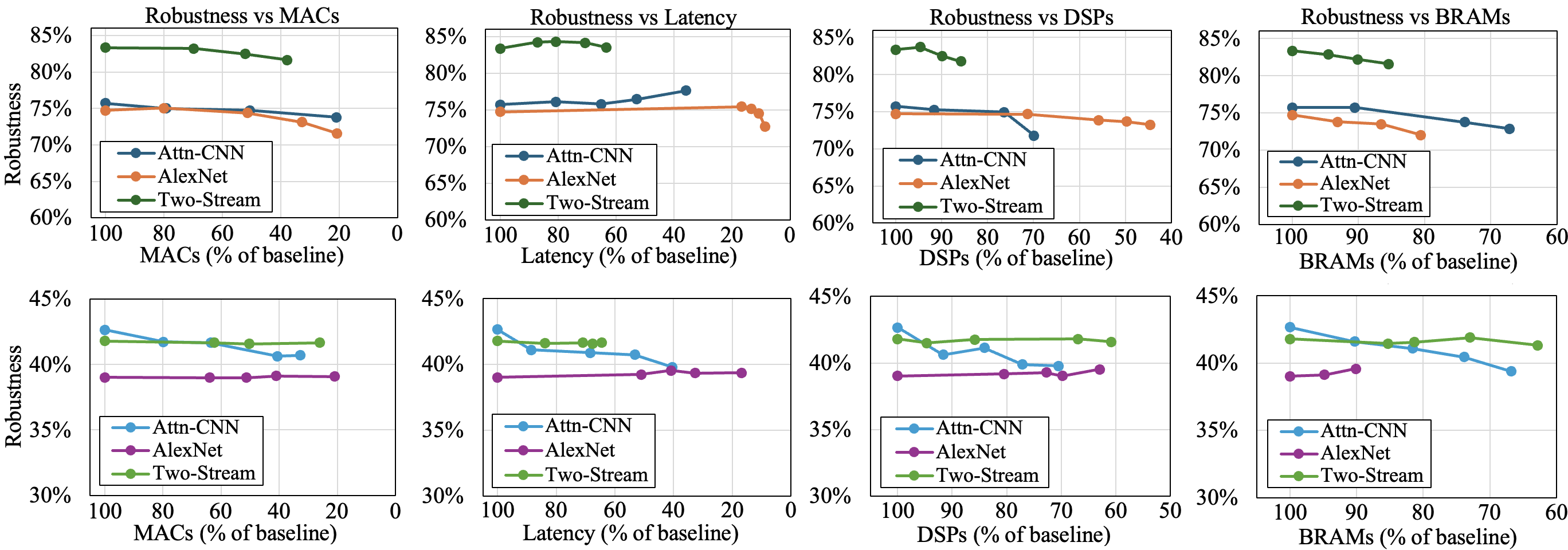}
  \caption{Robustness-efficiency trade-off for pruned and quantized models. (Top row: MSTAR, Bottom row: FUSAR-Ship)}
  \label{fig:pareto_mstar}
\end{figure}

\subsection{Robustness-Efficiency Trade-off Evaluation}

We evaluate the trade-off between robustness and inference efficiency by applying our hardware-guided pruning under four user-defined optimization objectives: MACs, latency, DSP usage, and BRAM usage. Figure~\ref{fig:pareto_mstar} presents the results for all models across the two datasets, which are fine-tuned, quantized, and evaluated on FPGA. 
Our pruning algorithm significantly reduces the MACs, latency, and resource requirements while preserving robustness. For example, the Attn-CNN optimized for latency on MSTAR, reduces inference latency to 36\% of the baseline latency while achieving 77.61\% robustness, which is even better than the unpruned version. 
AlexNet on MSTAR demonstrates significant efficiency gains, reducing latency to just 10\% of the baseline latency with only 0.25\% drop in robustness. The Two-Stream model on MSTAR maintains robustness above 81\% in all cases. We observe similar robustness-efficiency trends across other optimization objectives (MACs, DSPs, and BRAMs) demonstrating the effectiveness of our hardware-guided pruning approach.
The FUSAR-Ship results demonstrate similar patterns in the robustness-efficiency trade-offs, though the baseline robustness (and accuracy) levels are notably lower due to the challenging nature of the dataset (e.g., severe class imbalance). However, our pruning framework maintains robustness effectively across different compression levels.
In all cases, our pruning algorithm generates a diverse set of Pareto-optimal models, allowing users to select trade-offs that best fit their preference.





\subsection{Portability of the Hardware Design}
To validate the resource-adaptive nature of the proposed accelerator architecture, we evaluate the design on the ZCU104, a resource-constrained FPGA platform. The ZCU104 is comparable to state-of-the-art space-grade FPGAs, such as the Kintex Ultrascale XQRKU60, in terms of technology and available hardware resources. Table~\ref{tab:zcu_kintex} presents a detailed comparison between the ZCU104 and the Kintex Ultrascale XQRKU60. As shown, the ZCU104 shares similar on-chip memory capacity and follows a comparable architectural class in terms of logic and DSP resources, making it a practical prototyping platform for space-oriented deployments. Motivated by on-board SAR processing scenarios where resource availability is significantly more constrained, we implement our design on the ZCU104 to demonstrate portability across different resource budgets. In this setting, the accelerator is instantiated using the temporal resource-reuse architecture described in Section~\ref{sec:hw_design}, enabled by our parameterized HLS templates. We set the number of processing elements ($N_{\text{pe}}$) to 8, with an unroll factor of 8 applied within each compute unit, and synthesize the design to operate at  280~MHz. This implementation validates that the same model-hardware co-design framework can be deployed on smaller FPGA platforms without modifying the overall mapping flow.

\begin{table}[ht]
\centering
\caption{Specifications of the state-of-the-art space-grade FPGA and the target FPGA}
\begin{adjustbox}{max width=0.65\textwidth}
\begin{tabular}{cccc}
\toprule
\textbf{Platform} & \begin{tabular}[|c|]{@{}c@{}} \textbf{Space-grade FPGA}\\ Kintex Ultrascale XQRKU60 \end{tabular} & \begin{tabular}[|c|]{@{}c@{}} \textbf{Target FPGA} \\ ZCU104 \end{tabular}\\
\midrule
Technology & TSMC 20 nm & TSMC 16 nm  \\ \midrule
System Logic cells (K) & 726 & 504 \\ \midrule
DSP & 2760 & 1728 \\ \midrule
On-chip Memory (Mb) & 38 & 38 \\ \bottomrule
\end{tabular}
\end{adjustbox}
\label{tab:zcu_kintex}
\end{table}

\begin{table}[!ht]
\centering
\caption{Robustness-efficiency trade-off for pruned and quantized models on the resource-constrained FPGA}
\begin{adjustbox}{max width=0.95\textwidth}
\renewcommand{\arraystretch}{1.15}
\begin{tabular}{c|c|cccc|ccc}
\Xhline{0.8pt}
 \textbf{Model} & \makecell{\textbf{Pruning} \\ \textbf{Objective}} & \makecell{\textbf{Accuracy} \\ \textbf{(\%)}} & \makecell{\textbf{Robustness} \\ \textbf{(\%)}} & \textbf{MACs} & \makecell{\textbf{Model Size} \\ \textbf{(MB)}} & \makecell{\textbf{Latency} \\ \textbf{(ms)}} & \textbf{DSP} & \textbf{BRAM} \\
\Xhline{0.8pt}

\multirow{4}{*}{\textbf{Attn-CNN}} 
  & No Pruning & 97.53 & 75.71 & $5.85\times10^8$ & 0.72  & 17.38  & 882 (51.04\%) & 503 (80.61\%) \\ 
  \cline{2-9}
  & MACs & 97.48 & 73.77 & $1.23\times10^8$ & 0.54  & 5.00  & 882 (51.04\%) & 321 (51.44\%) \\ 
  & Latency & 97.15  & 70.76  & $0.80\times10^8$ & 0.48  & 3.33  & 882 (51.04\%) & 299 (47.92\%) \\ 
  & BRAM & 96.74 & 72.87 & $0.99\times10^8$ & 0.50  & 3.97 & 882 (51.04\%) & 299 (47.92\%) \\ 
\Xhline{0.8pt}

\multirow{4}{*}{\textbf{AlexNet}} 
  & No Pruning & 92.02 & 74.72 & $2.36\times10^8$ & 57.03  & 2.55  & 882 (51.04\%) & 529 (84.78\%) \\ 
  \cline{2-9}
  & MACs & 92.41 & 73.11 & $0.77\times10^8$ & 12.50  & 0.82  & 882 (51.04\%) & 404 (64.74\%) \\ 
  & Latency & 91.63  & 71.59  & $0.49\times10^8$ & 9.85  & 0.49  & 882 (51.04\%) & 377 (60.42\%) \\ 
  & BRAM & 92.91  & 74.47 & $1.04\times10^8$ & 9.20   & 1.29  & 882 (51.04\%) & 334 (53.53\%) \\ 
\Xhline{0.8pt}
\end{tabular}
\end{adjustbox}
\label{tab:zcu_prune}
\end{table}

Table~\ref{tab:zcu_prune} presents the robustness-efficiency trade-offs of pruned and quantized models implemented on the target resource-constrained FPGA. Compared to the unpruned baselines, all pruning objectives (MACs, latency, and BRAM) significantly reduce inference latency and memory footprint while incurring minimal drop in robustness. For example, the latency-optimized AlexNet reduces inference time from $2.55$ ms to $0.49$ ms with only a moderate robustness drop. Similarly, Attn-CNN achieves substantial latency reduction from $17.38$ ms to $3.33$ ms under latency-optimized pruning. As expected in the temporal resource-reuse design, DSP utilization remains largely fixed due to the predefined maximum parallelism. In contrast, BRAM utilization shows minor variations across pruned models, as smaller channel dimensions reduce the maximum required buffer sizes. These results demonstrate that our parameterized template design effectively adapts to different pruning configurations while satisfying strict resource constraints.

\begin{table}[!ht]
\centering
\caption{Comparison against existing FPGA accelerators}
\begin{adjustbox}{max width=0.8\textwidth}
\renewcommand{\arraystretch}{1.05}
\begin{tabular}{l|c c|c c}
\Xhline{0.8pt}
 & \textbf{SMART~\cite{wickramasinghe2025cnn-att4}} & \textbf{This Work} & \textbf{fpgaConvNet~\cite{venieris2018fpgaconvnet}} & \textbf{This Work} \\
\Xhline{0.8pt}
Platform & Alveo U280 & Alveo U280 & ZCU7045 & ZCU104 \\
Model & Attn-CNN & Attn-CNN & AlexNet & AlexNet \\
Frequency (MHz) & 260 & 260 & 125 & 125 \\
Latency (ms) & 0.14 & 0.32 & 8.22 & 0.94 \\
Energy Efficiency (mJ/image) & 2.42 & 7.36 & 32.88 & 2.82 \\
\Xhline{0.8pt}
\end{tabular}
\end{adjustbox}
\label{tab:fpga_comparison}
\end{table}

Table~\ref{tab:fpga_comparison} compares our implementation against existing FPGA-based SAR ATR and CNN accelerators. We evaluate our designs at the same clock frequencies as the respective FPGA accelerator baselines. On the ZCU104 platform, our design achieves $0.94$ ms latency and $2.82$ mJ/image energy efficiency for AlexNet, outperforming fpgaConvNet on a comparable embedded-class FPGA. While our design exhibits marginally higher latency and lower energy efficiency compared to SMART, this difference stems from architectural design trade-offs. SMART is optimized for aggressive parallelization, whereas our design adopts controlled parallelism through channel-aware PE allocation to ensure portability to resource-constrained platforms. By constraining the maximum number of instantiated PEs to maintain deployability across resource-constrained devices, we trade peak throughput on large FPGAs for broader applicability and consistent mapping across platforms. Furthermore, the architecture is designed such that reductions in channel dimensions directly translate into latency and memory savings, ensuring pruning gains are effectively realized on hardware. Overall, these results highlight that our framework enables robust SAR ATR deployment across both high-performance and resource-limited FPGA platforms.

\subsection{Effectiveness of Hardware-Guided Pruning}

\begin{figure}[!ht]
  \centering
    \begin{subfigure}[b]{0.33\textwidth}
    \centering
    \includegraphics[height=2.8cm]{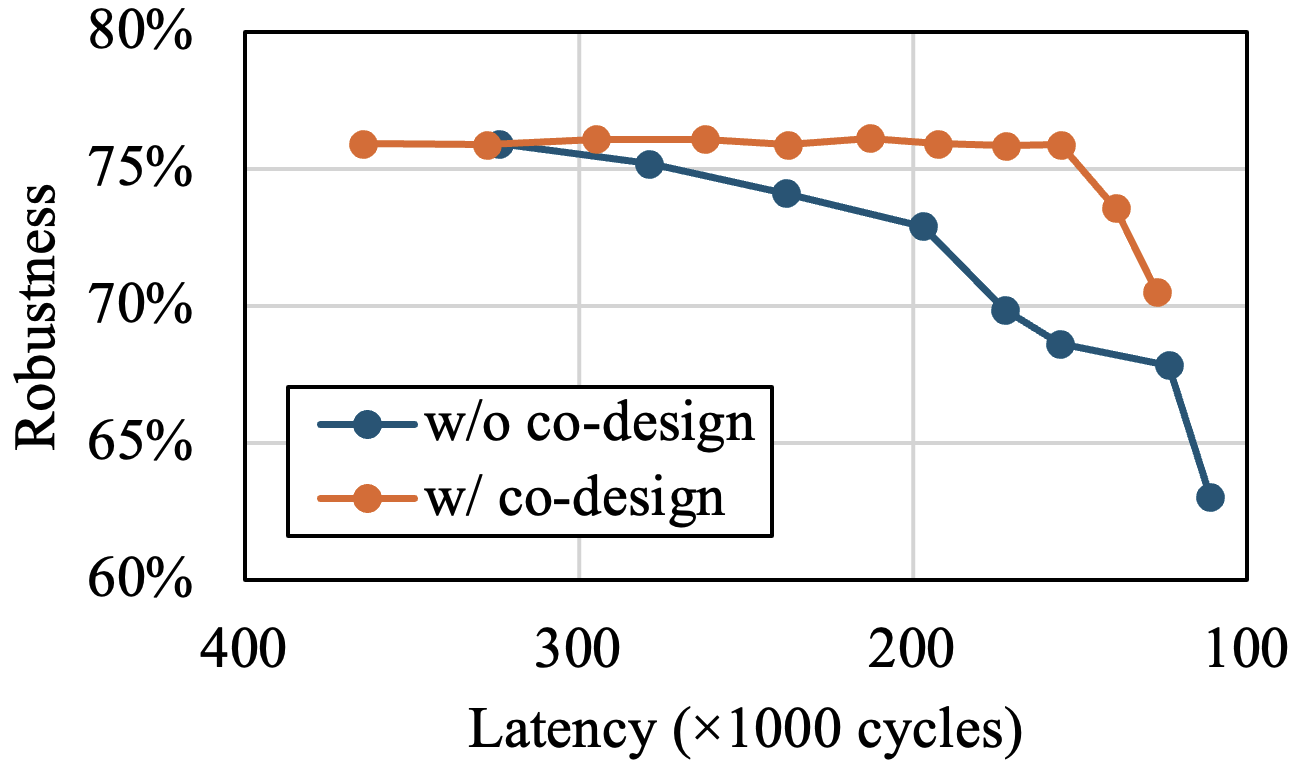}
    \label{fig:asm1}
    \end{subfigure}
    \begin{subfigure}[b]{0.31\textwidth}
    \centering
    \includegraphics[height=2.8cm]{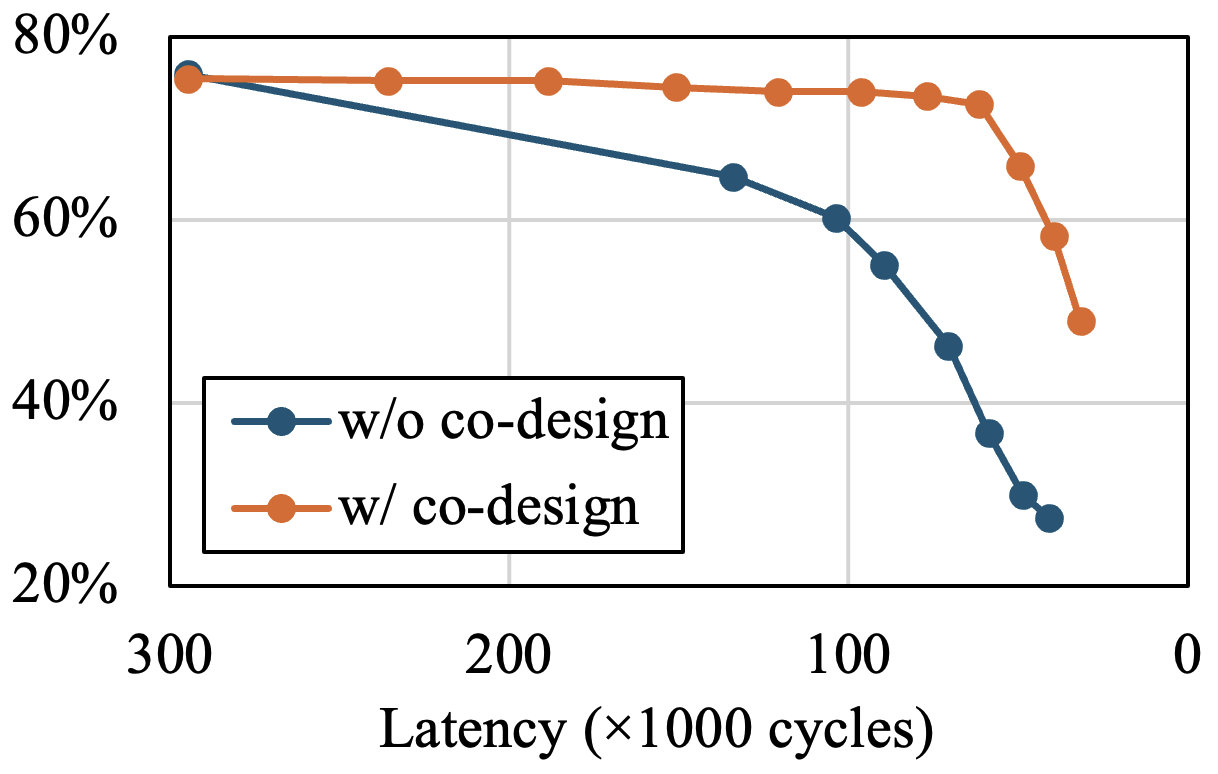}
    \label{fig:asm2}
    \end{subfigure}
    \begin{subfigure}[b]{0.31\textwidth}
    \centering
    \includegraphics[height=2.8cm]{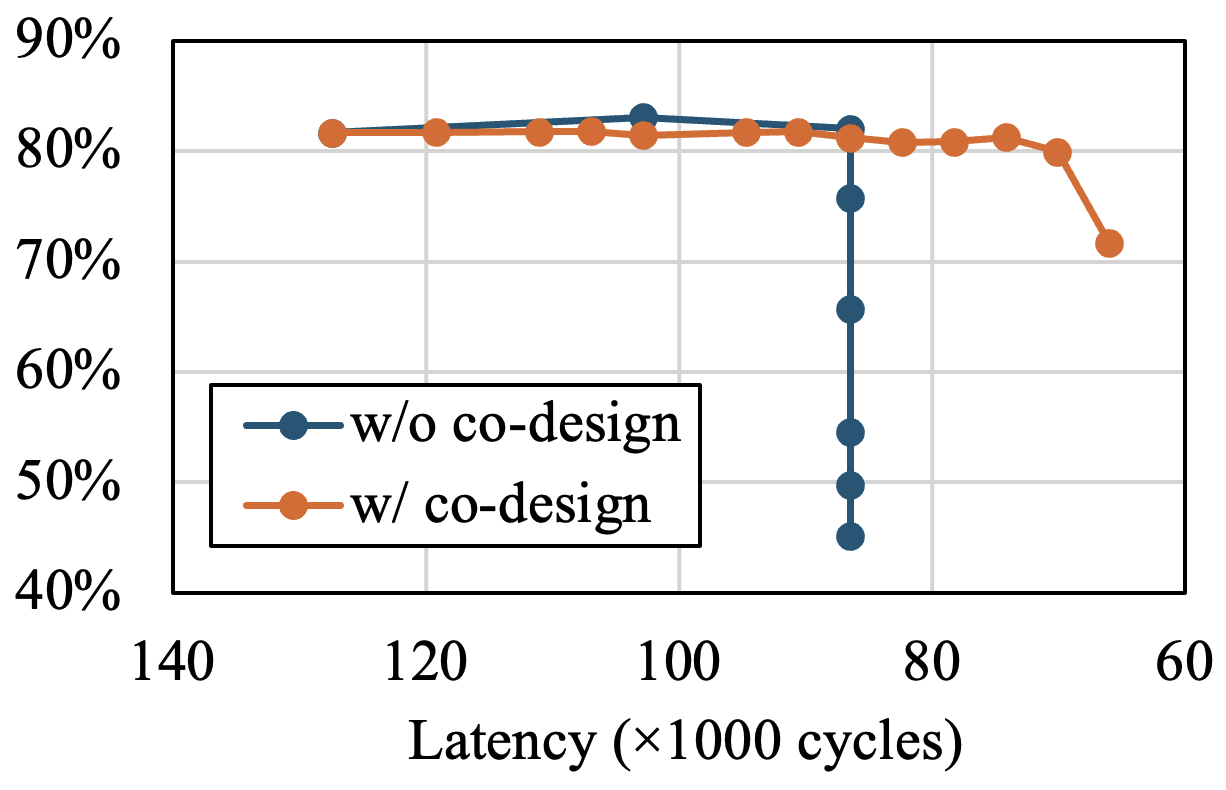}
    \label{fig:asm3}
    \end{subfigure}
  \caption{Robustness-latency trade-offs achieved by pruned models with and without guidance from hardware performance modeling. Experiments use MSTAR dataset. (Left: Attn-CNN, Middle: AlexNet, Right: Two-Stream)}
  \label{fig:co-design}
\end{figure}

To demonstrate the contribution of our co-design framework, we compare the result of model pruning under two settings: (1) with model-hardware co-design, where pruning decisions are guided by the performance model derived from the hardware design, and (2) without co-design, where channels are selected to prune based on saliency scores only. Figure~\ref{fig:co-design} shows the robustness-latency trade-offs of pruned models trained on the MSTAR dataset using three model architectures. To ensure fair comparisons, all experiments have no further fine-tuning to avoid potential randomness. For the version using the performance model, we use latency as the optimization objective. The results show that all models pruned with the guidance of the hardware performance model almost consistently achieves better robustness for the same latency. The gap between their robustness is particularly high when the model is pruned more aggressively. These results confirm that the hardware performance model is essential to effectively improve inference efficiency while preserving adversarial robustness, and the results highlight the critical role of model-hardware co-design in our framework. 

\begin{figure} [ht!]
\centerline{\includegraphics[width=0.85\linewidth]{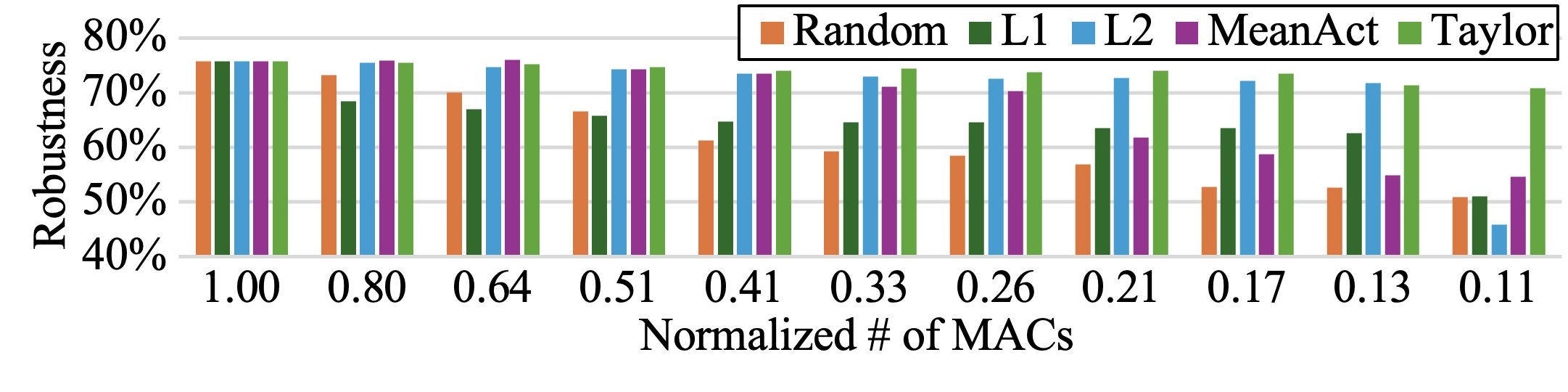}}
\caption{Robustness using different saliency functions.}
\label{fig:saliency}
\end{figure}

\subsection{Analysis of Saliency Functions}

We evaluate the four saliency score functions used in model pruning: $\ell_1$-norm, $\ell_2$-norm, mean activation, and first-order Taylor saliency, as defined in Section~\ref{sec:model-compression}. Besides, we also include a version that randomly choose channels to prune. The experiment is conducted on Attn-CNN with MACs as the pruning objective. Figure~\ref{fig:saliency} shows the robustness of the resulting models against the number of MACs normalized to the unpruned baseline. The results show that first-order Taylor saliency outperforms other functions in most cases, especially when the pruning is aggressive (e.g., with only 10\% MACs remaining). These findings support our choice of first-order Taylor saliency in the proposed pruning algorithm.




\subsection{Validation of Hardware Performance Model}
\label{sec:perf_model_val}

We validate the accuracy of our analytical hardware performance model by comparing its latency and resource usage estimates with measurements from the Vitis Analyzer. 
For latency of convolutional layers, we vary the number of channels to simulate different levels of pruning, and observe less than 2.5\% prediction error. For max-pooling layers, we vary the input feature map size from 8 to 128 and observe at most 2\% prediction error. 
In terms of resource usage, our estimation shows almost exact values in both BRAM and DSP predictions for convolutional layers across all pruning configurations and varying PEs. For max-pooling, DSP and BRAM estimates remain accurate, and deviate by at most 0.1\%. These small discrepancies are primarily due to pipeline overheads not explicitly captured in our performance model.

\section{Related Work}
\textbf{Model Compression:} Recent works on SAR ATR have explored a range of model compression strategies, including pruning, quantization, and knowledge distillation, to improve the deployability of CNNs on resource-constrained environments. For instance, \cite{wang2021boosting} combines structured pruning with channel attention mechanisms and knowledge distillation to significantly reduce model size. Similarly, \cite{chen2018slim} and \cite{babu2024optimizing} employ pruning, transfer learning, and lightweight architectures to create deployable models. However, these approaches typically explore compression methods in isolation and often ignore joint optimization of model and hardware.\\

\noindent \textbf{FPGA-based Acceleration:}
Recently, there has been growing interest in accelerating SAR ATR models on FPGA platforms to enable low-latency and energy-efficient deployment.
Several studies have explored customized FPGA accelerators for GNNs, ViTs, and CNNs tailored to SAR applications~\cite{gnn, bingyi_sar_atr_hbm, zhang2025model, wickramasinghe2024vtr, wickramasinghe2025cnn-att4, wickramasinghe2026model}.
Some works, such as OSCAR-RT~\cite{yang2022algorithm}, incorporate hardware-guided structured pruning for on-satellite ship detection. While these works advance hardware efficiency, they do not address adversarial robustness during the co-design process, a critical requirement for real-world SAR ATR systems. Existing works typically optimize a fixed network architecture and design corresponding hardware pipelines to maximize throughput and resource utilization. In contrast, our framework integrates an analytical hardware performance model directly into the pruning process, enabling systematic exploration of robustness-efficiency trade-offs. 
The resulting pruned models automatically configure parameterized HLS templates during synthesis, allowing the hardware architecture to adapt to varying channel dimensions without manual redesign. 
Existing FPGA-based SAR ATR accelerators focus exclusively on accuracy and efficiency, whereas our work incorporates adversarial robustness into the model-hardware co-design loop.
To the best of our knowledge, no prior work jointly optimizes adversarial robustness and hardware efficiency for SAR ATR.

\section{Conclusion and Future Work}

In this work, we presented a model-hardware co-design framework that jointly addresses adversarial robustness and inference efficiency for CNN-based SAR ATR on FPGA. 
The framework combines hardware-guided structured pruning of adversarially trained models with a parameterized FPGA accelerator design. An analytical performance model provides rapid hardware cost evaluation to guide pruning decisions, enabling systematic exploration of robustness-efficiency trade-offs. 
The framework employs parameterized modular compute engines that support both fully pipelined streaming and temporal resource-reuse architectures. We introduced an automated design generation flow to map selected pruned models into synthesizable HLS implementations. 
Experiments on MSTAR and FUSAR-Ship datasets using three SAR ATR CNNs demonstrate models up to $18.3\times$ smaller with $3.1\times$ fewer MACs while maintaining robustness within the predefined tolerance. 
The resulting FPGA implementations achieve up to $68.1\times$ ($6.4\times$) lower latency and up to $169.7\times$ ($33.2\times$) higher energy efficiency compared to CPU (GPU) baselines.
As future work, the robustness-aware pruning techniques developed in this work can be extended to other emerging deep learning models, including transformers, by adapting gradient-based saliency measures to identify and retain parameters most critical to maintaining adversarial robustness. 
These extensions can be applied across diverse safety-critical domains where adversarial robustness is essential. 
\begin{acks}
{This work is supported by the U.S. National Science Foundation (NSF) under grants ASUB00001630 and CCF-1919289/ SPX-2333009. Equipment support from AMD AECG is greatly appreciated.}
\end{acks}

\bibliographystyle{ACM-Reference-Format}
\bibliography{ref}










\end{document}